\documentclass{sig-alternate-2013}
\usepackage{algpseudocode}
\usepackage{amssymb}
\usepackage{amsmath}
\usepackage{graphicx}
\usepackage{setspace}
\usepackage{subfig}
\usepackage{xspace}
\usepackage{xcolor}
\usepackage{multirow}
\usepackage{paralist}
\usepackage{tabularx}
\usepackage{cite}

\usepackage{hyperref}
\hypersetup{
    bookmarks=true,         
    bookmarksopenlevel=2,
    unicode=false,          
    pdftoolbar=true,        
    pdfmenubar=true,        
    pdffitwindow=false,     
    pdfstartview={FitH},    
    pdftitle={},    
    pdfauthor={},     
    pdfsubject={},   
    pdfcreator={},   
    pdfproducer={}, 
    pdfkeywords={} {} {}, 
    pdfnewwindow=true,      
    colorlinks=false,       
    linkcolor=red,          
    citecolor=green,        
    filecolor=magenta,      
    urlcolor=cyan           
}
\usepackage{url}
\usepackage{breakurl}

\long\def\com#1{}

\long\def\xxx#1{}

\long\def\abbr#1#2{#1}      

\makeatletter
\renewcommand{\paragraph}{%
  \@startsection{paragraph}{4}%
  {\z@}{1ex \@plus 1ex \@minus .2ex}{-1em}%
  {\normalfont\normalsize\bfseries}%
}
\makeatother

\defaultleftmargin{\parindent}{}{}{}

\boilerplate={}
\confinfo={CCS'13, Nov 04-08 2013, Berlin, Germany}
\copyrightetc={ACM 978-1-4503-2477-9/13/11. \href{http://dx.doi.org/10.1145/2508859.2516740}{DOI 10.1145/2508859.2516740}}

\begin{document}

\hyphenation{An-o-nym-i-zer}
\hyphenation{an-o-nym-i-ty}
\hyphenation{an-o-nym-ous}
\hyphenation{pseu-do-nym}
\hyphenation{pseu-do-nym-i-ty}
\hyphenation{pseu-do-nym-ous}

\newcommand{\app}{Buddies\xspace}
\newcommand{\apps}{Buddies'\xspace}	

\newcommand{\anon}{an\-o\-nym\-i\-ty\xspace}
\newcommand{\pnym}{pseu\-do\-nym\xspace}
\newcommand{\possy}{pos\-si\-nym\-i\-ty\xspace}

\newcommand{\rosterset}{\mathbb{M}}
\newcommand{\userset}{\mathbb{O}}
\newcommand{\filtset}{\mathbb{P}}

\title{
	Hang With Your \app to Resist Intersection Attacks
}

\numberofauthors{1}
\author{
\alignauthor David Isaac Wolinsky, Ewa Syta, and Bryan Ford\\
       \affaddr{Yale University}\\
       \email{\{david.wolinsky,ewa.syta,bryan.ford\}@yale.edu}
}

\date{}

\maketitle

\begin{abstract}

Some anonymity schemes might in principle
protect users from pervasive network surveillance---%
but only if all messages are independent and unlinkable.
Users in practice often need {\em pseudonymity}---%
sending messages intentionally linkable
to each other but not to the sender---%
but pseudonymity in dynamic networks
exposes users to {\em intersection attacks}.
We present \app,
the first systematic design for intersection attack resistance
in practical anonymity systems.
\app groups users dynamically into {\em buddy sets},
controlling message transmission
to make buddies within a set behaviorally indistinguishable
under traffic analysis.
To manage the inevitable tradeoffs between anonymity guarantees
and communication responsiveness,
\app enables users to select
independent attack mitigation policies for each pseudonym.
Using trace-based simulations and a working prototype,
we find that \app can guarantee non-trivial anonymity set sizes
in realistic chat/microblogging scenarios,
for both short-lived and long-lived pseudonyms.

\end{abstract}




\section{Introduction}
\label{sec:intro}


Some anonymous communication techniques
promise security even against powerful adversaries
capable of pervasive network traffic analysis---%
{\em provided} all messages are fully independent of each other
and/or the set of participants never changes%
~\cite{chaum88dining,wolinsky12dissent,pfitzmann91isdn,berthold00cascade}.
Practical systems, however,
must tolerate {\em churn} in the set of online users,
and must support ongoing exchanges
that make messages {\em linkable} over time,
as with Mixminion nyms~\cite{danezis03mixminion}
or Tor sessions~\cite{dingledine04tor}.
By sending linkable messages in the presence of churn, however,
users can quickly lose anonymity
to statistical disclosure or intersection attacks~\cite{
raymond00traffic,kedogan02limits,danezis04statistical,wright08passive}.
Though this extensively studied attack vector
could apply in almost any realistic anonymous communication scenario,
no practical anonymity system we know of
offers active protection against such attacks.

As an example intended merely to illustrate
one possible scenario in this broad class of attacks,
suppose Alice writes a blog under a pseudonym to expose corruption
in her local city government.
Alice always connects to the blog server via Tor~\cite{dingledine04tor},
and never reveals personally identifying information
on her blog or to the server.
Carol, a corrupt city official mentioned in Alice's blog,
deduces from the blog's content that its owner is local,
and calls her friend Mallory,
a network administrator in the monopolistic local ISP.
Mallory cannot directly compromise Tor,
but she reads from Alice's blog the date and time
each blog entry was posted,
and she learns from the ISP's access logs
which customers were online and actively communicating
at each of those times.
While thousands of customers may be online at each posting time,
every customer {\em except} Alice
has a chance of being offline during {\em some} posting time,
and this chance exponentially approaches certainty
as Alice continues posting.
Mallory simply keeps monitoring until the {\em intersection}
of these online user sets narrows to one user, Alice.
We don't know if this precise attack has occurred,
but analogous intersections of hotel guest lists,
IP Addresses, and e-mail accounts revealed
the parties in the Petraeus/Broadwell scandal%
~\cite{soghoian12petraeus}.

\com{
\xxx{ good to start with a simple example,
	but this needs to be tightened up... -baf}
In one scenario, Alice, Bob, Carol, and Mallory agree
to use a secure chat room (akin to IRC)
built on top of Dissent~\cite{wolinsky12dissent},
a practical DC-net anonymity system.
Unbeknownst to Alice, Bob, and Carol, who genuinely desire anonymity,
Mallory has joined the chat room to identify the
perpetrator in some recent sensitive information leaks.
At random intervals throughout the next week,
the more information continues to leak,
when out of the blue, Bob disappears from the chat room,
perhaps his computer restarted due to a Windows' update mandatory reboot.
The message posting continues on uninhibited.
After Bob comes back,
Carol disappears,
perhaps a weak wireless signal.
Yet the message posting continues on uninhibited.
Throughout the poster assumed
the anonymity set size was at least 3;
however, Bob and Carol can be intersected out
leaving Alice as the likely source of the information leakage.
In general,
Alice, Bob, and Carol may represent many independent individuals
and Mallory may be either a member
or able to monitor network communication.
Regardless, if sufficient members have differing online times,
Mallory can eventually deanonymize members.
}

As a step toward addressing such risks
we present \app,
the first anonymous communication architecture
designed to protect users systematically
from long-term intersection attacks.
\app works by continuously maintaining
an anonymized database of participating users and their online status,
and uses this information to {\em simulate} intersection attacks
that a network-monitoring adversary might perform.
These simulations yield two relevant anonymity metrics
that \app reports continuously, as an indication
of potential vulnerability to intersection attack:
a {\em possibilistic} metric roughly measuring ``plausible deniability,''
and a more conservative {\em indistinguishability} metric
indicating vulnerability
to more powerful statistical disclosure attacks~\cite{danezis04statistical}.

Beyond just measuring vulnerability,
as in prior work on metrics~\cite{diaz02measuring,serjantov02metric}
and alternate forms of anonymity~\cite{hopper06on},
\app offers {\em active control}
over anonymity loss under intersection attack.
Users specify a policy for each \pnym that balances attack protection
against communication responsiveness and availability.
To enforce these policies,
a {\em policy module}
monitors and filters the set of users participating in each communication round,
sometimes forcing the system to behave as if certain online users
were actually offline.
Through this active control mechanism,
policies can enforce lower bounds on anonymity metrics,
preventing Alice from revealing herself to Mallory
by posting at the wrong time for example.
Policies can also reduce the {\em rate} of anonymity loss
to intersection attacks,
for example by tolerating anonymity set members
who are normally reliable and continuously online
but who lose connectivity for brief periods.
Finally, policies can adjust posting rates or periods,
enabling \app to aggregate all users coming online within a posting period
into larger anonymity sets.
If Alice sets her blog's posting period to once per day,
for example,
then \app can maintain Alice's anonymity
among all users who ``check in'' at least once a day---%
{\em any time} during each day---%
even if many users check in only briefly at widely varying times.

\apps architecture may be treated as an extension to
various existing anonymous communication schemes,
but is most well-suited to schemes already offering
measurable protection guarantees against traffic analyis,
such as MIX cascades~\cite{pfitzmann91isdn,berthold00cascade},
DC-nets~\cite{chaum88dining,sirer04eluding,wolinsky12dissent},
or verifiable shuffles~\cite{neff01verifiable,furukawa01efficient,brickell06efficient}.
We have built a working prototype of \app atop
Dissent~\cite{corrigangibbs10dissent,wolinsky12dissent,corrigangibbs13verif},
a recent anonymous communication system
that combines verifiable shuffle and DC-net techniques.
The prototype's design addresses several practical challenges:
to decentralize trust among independent servers,
to create and manage pseudonyms while maintaining their independence,
and to support user-selectable policies for each pseudonym.

To evaluate \apps practicality in realistic online communities,
we analyze IRC trace data under a \app simulator,
exploring questions such as how effective \apps anonymity metrics are,
how feasible it may be to maintain nontrivial anonymity sets
resistant to intersection attacks for extended periods,
and how effectively \app can limit loss of anonymity
while preserving usable levels of communication responsiveness and availability.
\xxx{
We find that ...
}

This paper's primary contributions are:
(a) the first \anon architecture
that systematically addresses intersection attacks;
(b) a modular, policy-based framework for both vulnerability monitoring
and active mitigation of anonymity loss via intersection attacks; and
(c) an evaluation of \apps practicality via a working prototype
and trace-based simulations reflecting realistic online communities.

Section~\ref{sec:model} of this paper outlines
\apps high-level model of operation
and the anonymity metrics we use.
Section~\ref{sec:policy} then explores
several useful attack mitigation policies in this model.
Section~\ref{sec:design} details challenges and approaches
to incorporating \app into practical anonymity systems, and
Section~\ref{sec:eval} experimentally evaluates
both our working \app prototype and trace-based simulations.
Section~\ref{sec:related} summarizes related work, and
Section~\ref{sec:conc} concludes.

\section{Buddies architecture}
\label{sec:model}

\begin{figure}[t]
\centering
\includegraphics[width=0.40\textwidth]{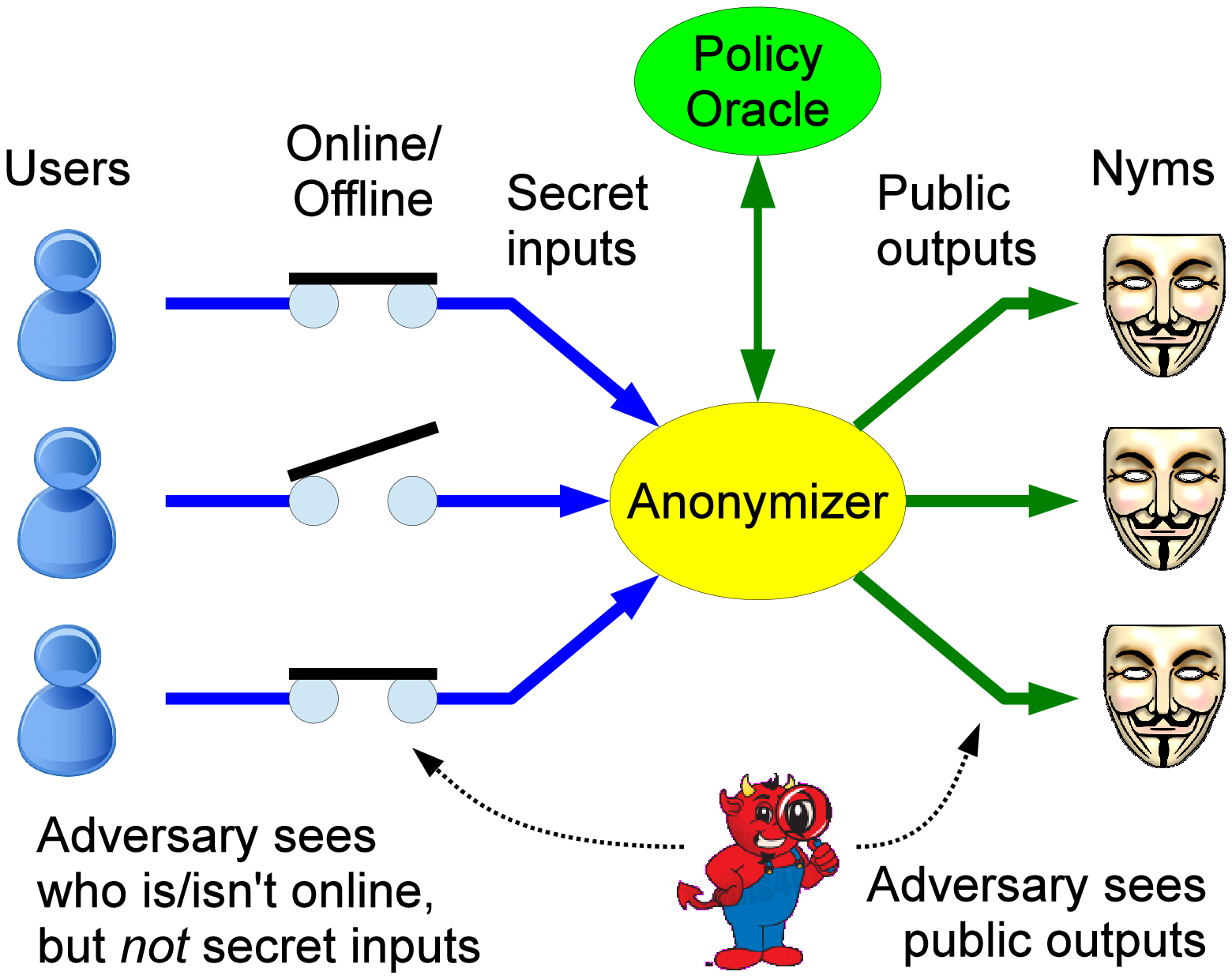}
\caption{Conceptual model of Buddies architecture}
\label{fig:arch-diagram}
\end{figure}

Figure~\ref{fig:arch-diagram}
shows a high-level conceptual model of the \app architecture.
Buddies assumes there is some set of {\em users},
each of whom has a secret (i.e., securely encrypted)
network communication path to a component
we call the {\em Anonymizer}.
For now we conceptually treat the Anonymizer
as a central, trusted ``black box,''
although later	
we will map this conceptually centralized component
to realistic anonymization systems that decentralized trust,
to avoid trusting any single physical component or administrative domain.

\apps model is inspired by anonymous blogging or IRC scenarios,
where users post messages to a public forum,
and users primarily desire sender anonymity~\cite{pfitzmann10terminology}.
While we expect \app to generalize
to two-way models and metrics~\cite{shmatikov06measuring},
we defer such extensions to future work.
Each \app user ``owns'' some number of {\em Nyms},
each representing a pseudonymous identity under which the owner may post:
e.g., an anonymous chat handle or blog.
Users may secretly submit messages to be posted to Nyms they own,
which the Anonymizer scrubs of identifying information and
publicly ``posts'' to that Nym.
To make various operational decisions,
the Anonymizer consults a {\em Policy Oracle}.
By design the Policy Oracle has no access to sensitive information,
such as who owns each Nym:
the Policy Oracle makes decisions based purely
on public information available to anyone.

We assume the network-monitoring adversary identifies users
by some network identifier or locator, such as IP address.
By monitoring these locators
the adversary can tell which users are online or offline
at any given moment,
and how much data they transmit or receive,
but {\em cannot} see the actual content of data communicated
between honest users and the Anonymizer.
These assumptions model an ISP-grade adversary
that can implement ``wholesale'' network-level monitoring of users
connected via that ISP.

\subsection{Overview of Operation}
\label{sec:model-operation}

In \apps conceptual architecture,
communication proceeds synchronously through a series of {\em rounds}.
The Anonymizer drives the operation of each round $i$, as follows:

{\bf 1. Registration:}
	At the start of round $i$ the Anonymizer
  updates the membership roster, $\rosterset_i$,
  to include members who may have recently joined.

{\bf 2. Nym creation:}
	The Anonymizer next creates and announces one ``fresh'' Nym $N_i$
	each round.
	For each new Nym,
	the Anonymizer chooses one User {\em uniformly at random}
	as the Nym's owner,
	keeping this ownership secret.
	A Nym's lifetime is in principle unlimited:
	over time users acquire fresh Nyms
	at random but ``statistically fair'' times.
	(We later address creation of larger ``batches'' of Nyms efficiently,
	so new users need not wait a long time before they can post.)

{\bf 3. Scheduling:}
	The Anonymizer consults the Policy Oracle to choose
	one Nym, $T_i$, for transmission in this round,
	from all Nyms in existence.
	The Policy Oracle also specifies
	the number of bits $B_i$ that the owner of Nym $T_i$ may post.
	(Scheduling multiple Nyms per round is a straightforward extension.)
	As the Policy Oracle can access only public information,
	scheduling cannot depend directly on
	which Users currently ``have messages to post.''
	Scheduling can depend on other factors, however,
	such as Nyms' lifetimes, recent usage,
	or the interest of other users as indicated in
	messages previously posted anonymously via other Nyms.

{\bf 4. Message submission:}
	The Anonymizer announces the scheduled Nym $T_i$
	and transmission length $B_i$ to the Users currently online.
	Each online user submits exactly $B_i$ secret bits to the Anonymizer.
	These secret bits may contain either ``real'' data,
	or a {\em null} message of $B_i$ zero bits, if
	the user has nothing useful to transmit at the moment.
	The bits sent from any user $j$ other than the owner of Nym $T_i$
	represent ``cover traffic'' necessary
	to hide the Nym-owner's message submission from traffic analysis.
	The Anonymizer forms an {\em online user set},
	$\userset_i \subseteq \rosterset_i$,
	consisting of the users who
	submitted a (real, null, or cover) message in round $i$.

{\bf 5. User filtering:}
	The Anonymizer now consults with the Policy Oracle,
	giving the Policy Oracle the set $\userset_i$ of online users---but
	{\em not} any message content or information about which, if any,
	of these users owns the Nym scheduled this round.
	The Policy Oracle returns
	a new, {\em filtered} user set $\filtset_i \subset \userset_i$,
	further constraining the set of online users whose submissions
	the Anonymizer will actually {\em accept} this round.

{\bf 6. Message posting:}
	If the owner of the scheduled Nym $T_i$
	is a member of $\filtset_i$---i.e., is online {\em and}
	was not filtered above---then
	the Anonymizer decrypts that user's secret message
	and posts it in association with Nym $T_i$.
	If the owner of $T_i$ is not in $\filtset_i$---either
	because the owner was not online
	or was filtered above---then
	the Anonymizer posts $B_i$ zero bits to Nym $T_i$:
	an output indistinguishable from a null message.

\subsection{Active Mitigation of Intersection Attacks}
\label{sec:model-active}

The user filtering step above (step 5)
serves as \apps primary ``control point''
through which to resist intersection attacks.
The Policy Oracle
uses publicly available information to {\em simulate} a virtual Adversary,
by continuously performing an ``intersection attack'' against each Nym.
At step 5 of each round $i$, the Policy Oracle
first forms an attack model for the scheduled Nym $T_i$,
based on prior history and the set $\userset_i$ of users online in this round.
The Policy Oracle computes one or more relevant anonymity metrics
as detailed further below,
and determines if action is required to limit or avoid anonymity loss
in this round.
If no action is required,
the Policy Oracle returns the unfiltered user set to the Anonymizer,
i.e., $\filtset_i = \userset_i$.
If action is required, however,
then the Policy Oracle can filter the user set
producing a $\filtset_i \subset \userset_i$,
thus preventing any user not in $\filtset_i$ from posting,
as if more users were offline
than are {\em actually} offline.

To illustrate how this filtering
enables the Policy Oracle to mitigate intersection attacks,
consider the following straw-man policy.
In step 5 of each round, the Policy Oracle
simply checks whether {\em all} known users are presently online,
i.e., whether $\userset_i = \rosterset_i$,
returning $\filtset_i = \rosterset_i$ if so,
and otherwise returning $\filtset_i = \emptyset$.
In effect, the Policy Oracle forbids the system from making progress---%
allowing {\em anyone} to post to {\em any} Nym---%
except when {\em all} users are online.
Since messages are posted only when all users are online,
the intersection of all {\em nonempty} rounds' user sets is $\rosterset_i$,
and the system preserves ``perfect'' anonymity
assuming the Anonymizer performs as required.
The tradeoff, of course, is effective system availability,
which would be unusable in most practical situations.
\xxx{	The addition of the non-static roster 
	makes the $\rosterset_i$ above technically incorrect,
	in a way that's nontrivial to fix.
	But probably not important enough to worry about now.}

The key technical challenge,
and a primary contribution of this paper,
is developing more nuanced methods of controlling
the user filtering step in each round.
By controlling these filtering choices,
we seek to maintain both measurable anonymity levels under intersection attack
{\em and} ``usable'' levels of availability,
under arguably realistic conditions.

There are situations
in which {\em no} active control mechanism will help.
If all users but one go offline permanently, for example,
the Policy Oracle has only two choices:
eventually allow the one remaining user to post,
giving up all anonymity under intersection attack;
or filter that user forever,
giving up availability completely.
Thus, we must set realistic expectations.
Section~\ref{sec:eval} experimentally investigates
the feasibility of resisting intersection attacks in IRC communities,
and tests possible control policies against feasibility metrics.

\apps architecture separates the Policy Oracle from the Anonymizer,
giving the Policy Oracle access {\em only} to ``public information''
we assume is known to everyone including the adversary,
eliminating the risk that policies
may ``accidentally'' compromise anonymity by leaking Nym ownership.
\abbr{}{
We can envision many potentially interesting policies
that may yield different filtering decisions at step 5 of each round
depending, for example, on the true owner of the scheduled Nym
is online in this round.
However, these type of policies tend to
leak anonymity rapidly in unexpected ways.
}
By architecturally disallowing the Policy Oracle
from having access to private information,
we avoid the need to analyze each policy carefully
for such ``side-channel'' anonymity leaks,
and instead can focus purely on the main question of how effectively
a policy mitigates intersection attacks
while maintaining usable levels of availability.

Another issue
is whether the information the Policy Oracle needs
to simulate the Adversary's intersection attacks---%
such as the set of users online in each round---%
{\em should} be considered ``public information.''
Although an ideal global adversary
would know this information anyway,
more realistic adversaries may be unable to monitor {\em all} users.
If \apps design ``hands out'' 
information that would otherwise be at least partially private---%
such as the IP addresses of all online users---%
we risk accidentally ``strengthening'' a weak adversary
into an effectively omniscient adversary.
In the important case of users' network identities
such as IP addresses,
our design mitigates this leak
by replacing IP addresses with anonymized {\em tags}
when reporting online user sets to the Policy Oracle,
as discussed in Section~\ref{sec:design-ident}.
However, whether \apps simulation-based architecture
may strengthen weak adversaries in other unexpected ways,
by making ``too much information'' public,
merits further study.

\subsection{Analyzing Intersection Attacks}
\label{sec:model-analysis}

While we do not attempt full formal analysis,
the simplicity of the conceptual model facilitates
straightforward informal analysis.
Our focus here is on a particular class of attacks,
namely what an adversary can learn
from users' online status over time
(the ``switches'' in Figure~\ref{fig:arch-diagram}).
The many other known attacks against practical anonymity systems
are important but out of this paper's scope.
We also claim no particular novelty in our analysis techniques or metrics;
our goal is merely to apply known
attacks~\cite{raymond00traffic,kedogan02limits,danezis04statistical}
and anonymity metrics~\cite{diaz02measuring,serjantov02metric}
to the \app model.

We assume the Anonymizer is trusted to ``do its job,''
keeping secret the linkage between Users and the Nyms they own.
We also assume honest Users---%
those users we care about protecting---%
do not ``give away'' their identities
or the relationships between their Nyms via the messages they post.

Under these conditions,
the Adversary obtains three potentially important
pieces of information in each round $i$:
(a) the set of online users $\userset_i$,
(b) the set $\filtset_i$ of online users who
	passed the Policy Oracle's filter in step 5, and
(c) the $B_i$ message bits that were posted to the scheduled Nym $T_i$.
An observation that will be key to \app's design is that
only (b) and (c) will {\em actually} prove relevant
to intersection attack analysis:
the adversary ultimately obtains no useful information
from knowing which users {\em were online} during a given round,
beyond what the adversary learns
from knowing which users {\em were online and unfiltered}.

Since we assume honest users do not ``give away'' their identities
in their message content,
we ultimately care only whether the message posted to Nym $T_i$
was null or non-null.
If a non-null message appeared for Nym $T_i$ in round $i$,
then the adversary infers that the owner of $T_i$ must be
a member of the filtered user set $\filtset_i$ in that round.
(If the owner of $T_i$ was online but filtered in round $i$,
then a null message would have appeared, as if the owner was offline.)

If a {\em null} message appears for $T_i$,
however,
the Anonymizer's design ensures that the adversary
cannot distinguish among the following three possibilities:
(1) the owner of Nym $T_i$ was offline,
(2) the owner was online but filtered, or
(3) the owner was online and unfiltered,
but had nothing useful to send
and thus intentionally posted a null message.

\com{
We now consider the adversary's viewpoint from two levels of analysis:
possibilistic and probabilistic.
}

\subsubsection{Possibilistic Anonymity Analysis}
\label{sec:model-poss}

To construct a simple {\em possibilistic} anonymity set $P_N$
for a given Nym $N$,
the adversary intersects the filtered user sets $\filtset_i$
across all rounds $i$ for which Nym $N$ was scheduled
{\em and} a non-null message appeared:
i.e., $P_N = \bigcap_i \{ O_i \mid T_i = N \mathrel{\land} m_i \ne 0 \}$.
Thus, $P_N$ represents the set of users
who might conceivably own Nym $N$,
consistent with the observed set of non-null messages that have appeared
for Nym $N$ up to any point in time.
Since the adversary cannot distinguish whether
the appearance of a null message means that $N$'s owner
was offline, filtered, or merely had nothing to send,
null-message rounds do not eliminate the possibility
that users offline during that round may own $N$,
so such rounds leave the possibilistic anonymity set $P_N$ unaffected.

We define the size of a Nym's possibilistic anonymity set, $|P_N|$,
as Nym $N$'s {\em possibilistic anonymity},
which for convenience we abbreviate as {\em possinymity}.
Although possinymity is only one of the many useful anonymity metrics
that have been proposed~\cite{kelly08towards,kelly09taxonomy},
and perhaps a simplistic one,
we feel it captures a useful measure of ``plausible deniability.''
If for example a user is dragged into court,
and the judge is shown network traces of a \app system
in which the accused is one of $|P_N|$ users
who {\em may in principle} have posted an offending message,
then a large possibilistic anonymity may help sow uncertainty
of the user's guilt.
We fully acknowledge the weaknesses of plausible deniability in general,
however,
especially in environments where ``innocent until proven guilty''
is not the operative principle.

\subsubsection{Probabilistic Anonymity Analysis}
\label{sec:model-prob}

While a simplistic adversary might stop at the above analysis,
a smarter adversary can probabilistically learn
not only from rounds in which non-null messages appeared,
but also from rounds in which only a null message appeared.

Suppose for example the adversary correctly surmises from past observation that,
in each round $i$, the owner of Nym $N$
will have no useful message to post
with some independent and uniformly random probability $p$.
In this case the user will ``pass'' by submitting a null message.
With probability $1-p$ the user will have a non-null message
and will try to post it---%
but this post attempt fails, yielding a null message anyway,
if the owner is offline or filtered in that round.

For simplicity assume there are two users $A$ and $B$,
the adversary observes exactly one round $i$,
this round results in a null message,
and $\filtset_i = \{ A \}$:
user $A$ participated but user $B$ did not.
The null output from round $i$ means one of two events occurred:
(a) $A$ owns $N$, but chose with probability $p$ not to post in round $i$; or
(b) $B$ owns $N$, and no message appeared independently of $p$
because $B \not\in \filtset_i$.
Because Nyms are assigned to users uniformly at random on creation,
the ``base'' probability that either user owns $N$ is 1/2.
The probability of the above events (a) and (b) occurring
{\em conditioned} on the observed history, however, is different.
To be precise,
$P[\textrm{(a)} \mid \textrm{(a)} \cup \textrm{(b)}] =
	P[\textrm{(a)} \cap (\textrm{(a)} \cup \textrm{(b)})]
		/ P[\textrm{(a)} \cup \textrm{(b)}] =
	(p/2) / P[\textrm{(a)} \cup \textrm{(b)}]$,
and 
$P[\textrm{(b)} \mid \textrm{(a)} \cup \textrm{(b)}] =
	(1/2) / P[\textrm{(a)} \cup \textrm{(b)}]$.
Since (a) and (b) are disjoint events,
$P[\textrm{(a)} \cup \textrm{(b)}] = P[\textrm{(a)}] + P[\textrm{(b)}]$, so
$P[\textrm{(a)}] = (p/2) / (p/2 + 1/2) = p/(p+1)$, and
$P[\textrm{(b)}] = (1/2) / (p/2 + 1/2) = 1/(p+1)$.

From the adversary's perspective,
observing one round in which no message appears for Nym $N$,
and in which $A$ participated but $B$ did not,
reduces the relative likelihood of $A$ being the owner by a factor of $p$.
Observing similar events across multiple rounds
exponentially increases the adversary's ``certainty'' 
of $B$ being the owner:
after $k$ such rounds, the likelihood of $A$
being the owner is only $p^k/(p^k+1)$.
\com{
In a ``worst case'' when $p=0$,
meaning that the owner of Nym $N$ {\em always} posts a non-null message
whenever he can---and the adversary guesses this fact---%
the adversary learns with certainty that $B$ owns $N$
after only one round in which $B$ was excluded and no message appeared.
}

\subsubsection{Indistinguishability Under Probabilistic Attack}

The above reasoning generalizes to many users,
varying probabilities of posting, etc.
Our focus is not on deepening such analysis, however,
a goal admirably addressed
in prior work~\cite{diaz02measuring,serjantov02metric}.
Instead, we wish to achieve measurable resistance
to {\em unknown} probabilistic attacks:
we do not know the probabilities with which users
will attempt to post in particular rounds,
how well the unknown attacker may be able to predict
when the owner of a given Nym will post, etc.

Instead of relying on the relevance of
any {\em particular} probabilistic analysis---%
which may break each time a known attack is refined---%
\app relies on an {\em indistinguishability} principle
that applies to all attacks of this class.
If two users $A$ and $B$
have exhibited {\em identical} histories with respect to
inclusion in each round's filtered user set $\filtset_i$,
across {\em all} rounds $i$ in which a Nym $N$ was scheduled so far,
then under any probabilistic analysis of the above form
the adversary must assign identical probabilities to $A$ and $B$
owning Nym $N$.
That is, if for every round $i$,
it holds that $(A \in \filtset_i) \iff (B \in \filtset_i)$,
then users $A$ and $B$ are {\em probabilistically indistinguishable}
from each other, hence equally likely to own Nym $N$.

For any user $A$ and Nym $N$,
we define $A$'s {\em buddy set} $B_N(A)$ 
as the set of users probabilistically indistinguishable from $A$,
including $A$ itself,
with respect to potential ownership of Nym $N$.
If $n$ users are probabilistically indistinguishable from $A$,
then under the attacker's analysis
each such user in $B_N(A)$ has an individual probability
no greater than $1/|B_N(A)|$ of being the owner of $N$.
Intuitively, buddy-sets form equivalence classes of users
who ``hang together'' against probabilistic intersection attacks---%
so that individual buddies do not ``hang separately.''

We next define a second anonymity metric,
{\em indistinguishability set size}, or {\em indinymity} for short,
as the size of the {\em smallest} buddy-set for a given Nym $N$.
Since we do not know how a real attacker
will actually assign probabilities to users,
indinymity represents the minimum level of anonymity
a member of {\em any} buddy set can expect to retain,
even if the adversary correctly intersects the owner's anonymity set
down to the members of that buddy set.
Thus, the attacker cannot (correctly)
assign a probability greater than $1/|B_N|$ to {\em any} user---%
including, but not limited to, the owner of $N$.

One might argue that we ``mainly'' care
about the buddy set containing the true owner of $N$,
not about other buddy sets not containing the owner.
A counter-argument, however,
is that a particular observation history
might make some other buddy set
falsely appear to the adversary as ``more likely'' to own $N$.
In this case, we may well care how much protection
the innocent members of that ``unlucky'' buddy set
have against being ``falsely accused'' of owning $N$.
Thus, to ensure that {\em all} users have the ``strength in numbers''
of being indistinguishable in a crowd of at least $n$ users,
regardless of the adversary's probabilistic reasoning,
we must ensure that {\em all} buddy sets
have size at least $n$.

\section{Attack Mitigation Policies}
\label{sec:policy}

Based on the above architecture,
we now explore possible intersection attack mitigation policies.
We make no claim that these are the ``right'' policies,
merely a starting point for ongoing refinement.
Two key benefits of \apps architecture, however,
are to modularize these policies into replaceable components
independent of the rest of the anonymous communication system
so they can be further evolved easily,
and to ensure by system construction
that policies cannot leak sensitive information
{\em other than} by failing to protect adequately against intersection attacks.

We first explore policies for maintaining possinymity,
then policies to enforce a lower bound on indinymity.
An important caveat with any \anon metric
is that \app cannot guarantee that {\em measured} \anon
necessarily represents {\em useful} \anon,
if for example an attacker can compromise many users
or create many Sybil identities~\cite{douceur02sybil}.
Section~\ref{sec:malicious} discusses these issues in more detail.

\subsection{Maximizing Possinymity}
\label{sec:policy-poss}

The possinymity metric defined in Section~\ref{sec:model-poss}
considers only rounds in which non-null messages appear for some Nym $N$,
intersecting the filtered user sets across all such rounds
to determine $N$'s possinymity set $P_N$.
We consider several relevant goals:
maintaining a minimum possinymity level,
mitigating the {\em rate} of possinymity loss,
or both.

\paragraph{Maintaining a Possinymity Threshold}
Suppose a dissident, posting anonymously
in a public chat room under a Nym $N$,
wishes to maintain ``plausible deniability''
by ensuring that $|P_N| \ge 100$ throughout the con\-ver\-sa\-tion---%
and would rather be abruptly disconnected from the con\-ver\-sa\-tion
(or have Nym $N$ effectively ``squelched'')
than risk $|P_N|$ going below this threshold.
As a straightforward policy for this case,
at step 5 of each round $i$,
the Policy Oracle computes the new possinymity that $N$ would have
if $\userset_i$
is intersected with $N$'s ``running'' possinymity set from the prior round.
The Policy Oracle returns $\filtset_i = \userset_i$
if the new possinymity remains above threshold,
or $\filtset_i = \emptyset$ otherwise.

In practice, the effect is that $N$'s possinymity
starts out at an initial maximum of
the total set of users online when the user first posts via $N$,
then decreases down to (but not below) the possinymity threshold
as other users go offline, either temporarily or permanently.
This policy has the advantage of not reducing the usability of $N$,
or artificially delaying the time at which the user's posts appear,
as long as the possinymity set remains above-threshold.
Once $N$'s possinymity set reaches the set threshold, however,
all of the users remaining in this set become {\em critical},
in that $N$ becomes unusable for posting
once {\em any} remaining member goes offline.
In the ``dissident scenario'' we envision this event
might be the user's signal to move to a new network location:
e.g., get a fresh IP address at a different Internet cafe.

\paragraph{Limiting Possinymity Loss Rate}
An alternative, or complementary, goal is to reduce
the rate at which $N$'s possinymity decreases.
In realistic scenarios,
as our trace data in Section~\ref{sec:eval} illustrates,
clients often get delayed or disconnected temporarily
but return soon thereafter.
Thus, a more refined policy might temporarily
halt all posting for Nym $N$---by returning $\filtset_i = \emptyset$---%
when members of $N$'s current possinymity set go offline,
in hopes that the missing members will soon return.
To ensure progress and get $N$ ``unstuck''
if members remain offline for a sufficient number of rounds,
however, the policy eliminates these persistently offline members
from $N$'s permanent possinymity set by returning
a smaller (but nonempty) $\filtset_i$.
Such a policy may ``filter out'' possinymity set losses
due to otherwise-reliable users going offline briefly,
at the cost of
delaying a user's posting to Nym $N$ for a few rounds
if some current possinymity set member goes offline permanently.
Of course, a loss rate limiting policy may readily be combined
with a threshold-maintaining policy of the form above.
A further refinement of this combination might be to
increase the loss limiting policy's ``tolerance''---%
number of rounds a user may remain offline before being eliminated---%
as the Nym's possinymity set size falls
to approach the user's specified lower bound.
Such a policy in essence trades more {\em temporary} unavailability
for greater total Nym longevity.

\paragraph{Users Worth Waiting For}

While the above simple variants suggest starting points,
we envision many ways to refine policies further,
for example by recognizing that a user's record of past reliability
is often a predictor of future reliability.
To maximize a Nym $N$'s possinymity and minimize anonymity loss rate,
while also limiting delays caused by {\em unreliable} users,
we may wish to consider some members of $N$'s current possinymity set
to be ``more valuable'' than others:
e.g., users who have remained online and participating reliably
for a long period with at most a few brief offline periods.
In particular, a policy might apply an offline-time threshold
as discussed above to limit loss rate,
but apply {\em different} thresholds to different members
of $N$'s current possinymity set,
giving longer, more generous thresholds to more ``valued'' users.

The Policy Oracle can build up reliability information about users
starting when those users first appear---%
not just when a particular {\em pseudonym} of interest is created.
Thus, a policy that \app applies to a particular Nym $N$
can benefit from user history state that the Policy Oracle may have built up
since long before $N$ was created.

\subsection{Guaranteeing Minimum Indinymity}
\label{sec:policy-ind}

The possinymity metric
considers intersection attacks only across rounds
in which non-null messages appear,
but in realistic chat or blogging scenarios
a user's posts will be interspersed with idle periods
during which {\em no} message appears.
A smarter adversary can use the predictive techniques
discussed in Section~\ref{sec:model-prob}
to glean probabilistic information from such rounds.
We therefore wish to guarantee users
some level of indinymity, even under probabilistic attacks.

\paragraph{Forming and Enforcing Buddy-Sets}
To guarantee a Nym $N$ will have a minimum indinymity of some value $K$,
the Policy Oracle must ensure that all of $N$'s buddy-sets---%
subsets of users whose members exhibit {\em identical}
online behavior across rounds after user filtering---%
are all of size at least $K$.
As a straw-man approach to buddy-set formation,
on the creation of Nym $N$,
the Policy Oracle could divide $N$'s initial user roster $\rosterset$
into $\lfloor |\rosterset|/K \rfloor$ arbitrary buddy-sets ``up-front,''
each containing at least $K$ users.
At step 5 of each round $i$,
for each buddy-set $B$ containing
{\em any} offline user $u \not\in \userset_i$,
the Policy Oracle removes {\em all} members of $u$'s buddy set $B$
from its filtered user set $\filtset_i$.
The Policy Oracle effectively {\em forces} members of each buddy set
to come online or go offline ``in unison,''
keeping them permanently indistinguishable
under probabilistic intersection attack
(even if the adversary can distinguish one buddy set from another).
This straw-man policy is likely to yield poor availability, however,
as it prevents $N$'s owner from posting whenever {\em any}
member of the owner's buddy-set is offline,
making $N$ unusable in the fairly likely event
that any member of this buddy set is unreliable or disappears permanently.

\paragraph{Lazy Buddy-Set Formation}
As a first refinement, therefore,
the Policy Oracle can delay buddy-set decisions
until users actually go offline.
At creation, a Nym $N$ starts with one large buddy-set
containing its entire user roster $\rosterset$.
In the first round $i$ in which member(s) of $\rosterset$ go offline,
the Policy Oracle might first delay {\em all} posting for Nym $N$
by returning $\filtset_i = \emptyset$,
in hopes the missing member(s) will return online soon,
as discussed above.
Once the Policy Oracle ``gives up'' on one or more members,
however,
it splits $N$'s current buddy-sets into two,
isolating all persistently offline members into one of the resulting buddy-sets.
By delaying the decision of how to split buddy-sets,
the Policy Oracle guarantees that after the split-point,
the buddy-set containing only online members will
have a chance to ``make progress''---%
at least until more users go offline for sufficiently long
to force another buddy set split.

After a split,
each of of the resulting buddy-sets must be at least of size $K$
to maintain an indinymity lower bound.
If fewer than $K$ total users
are actually offline ($|\rosterset-\userset_i|<K$),
then the Policy Oracle must ``sacrifice'' a few online users,
placing them in the offline users' buddy-set.
\com{to bring its size up to $K$.}
Otherwise, if for example a single offline user forcing a split
is actually the owner of Nym $N$\com{---%
quite possible since the Policy Oracle does not know
which user owns $N$---%
},
then placing the offline user in a buddy set of size 1
would quickly reveal to a probabilistic attacker
that the now-offline user owned the pseudonym,
once the attacker notices that posts have stopped appearing on Nym $N$.
By ``sacrificing'' $K-1$ additional users at the split point,
even if the attacker infers
from the absence of posts that $N$'s owner
is in the now-offline buddy-set,
he cannot tell whether the owner is the user who {\em caused} the split,
or is one of those sacrificed and {\em forced} offline to ``keep him company.''
If the offline users in a buddy set eventually return online,
then the whole buddy set rejoins in unison,
making the Nym usable again if the owner was a member of this buddy set.

\paragraph{Choosing Whom to Sacrifice}
\label{sec:model-buddy-sacrifice}
When the Policy Oracle must ``sacrifice'' online users
to pad an offline buddy set to size $K$,
an important issue is how to choose which online users to sacrifice.
We investigated two classes of sacrificial policies:
random, and least reliable users.
Random choice clusters users into buddy sets
regardless of reliability,
which is simple but risks sacrificing reliable users
by mixing them into buddy sets containing unreliable
or permanently offline users.
Alternatively, the Policy Oracle might first sacrifice users
with the weakest historical record of reliability:
e.g., users who arrived recently or
exhibited long offline periods.
By this heuristic we hope to retain the most reliable users
in the buddy set that remains online immediately after the split---%
though this buddy set may split further
if more nodes go offline in the future.

We expect reliability-sensitive policies
to maximize a Nym's effective lifetime,
{\em provided} the Nym's true owner is one of the more reliable users
and does not get sacrificed into an unreliable
or permanently offline buddy set.
A short-lived or unreliable user cannot expect his Nyms to be long-lived
in any case:
a long-lived Nym must have a ``base'' of reliable, long-lived users
to maintain anonymity under intersection attack,
and the Nym's owner must obviously {\em be} a member
of the long-lived anonymity set he wishes to ``hide in.''


\subsection{Varying Policies and Nym Independence}
\label{sec:policy-varying}

So far we have assumed the Policy Oracle
enforces a ``global policy'' on all Nyms,
but this would limit flexibility and
perhaps unrealistically require all users to ``agree on'' one policy.
Instead, \app allows each Nym to have a separate policy---%
e.g., different lower bounds
and anonymity loss mitigation tradeoffs---%
chosen by the Nym's owner.

Since intersection attacks are by definition not an issue
until a Nym $N$ has been in use for more than one round,
$N$'s owner specifies the policy parameters for a Nym $N$
in its first post to $N$.
The set of users online in this first message round,
in which $N$'s policy is set,
forms $N$'s initial user roster $\rosterset$,
which is also by definition the maximum anonymity set $N$ can ever achieve
under intersection attack.
In subsequent rounds in which Nym $N$ is scheduled,
the announced policy for $N$ determines the Policy Oracle's behavior
in filtering $N$'s user sets to mitigate intersection attacks.

Each Nym's policy is thus independent of other Nyms---%
including other Nyms owned by the same user.
This policy independence,
and the correctness of Section~\ref{sec:model}'s analysis,
depend on the assumption made in Section~\ref{sec:model-operation}
that the Anonymizer assigns Nyms to users
{\em uniformly at random and independent of all other Nyms}.
Otherwise, the choices the Policy Oracle makes on behalf of one Nym
might well leak information about other Nyms.
This leads to some specific design challenges
addressed below in Section~\ref{sec:design-indep}.

To illustrate the importance of independent Nym assignment,
suppose there are two users,
to whom the Anonymizer {\em non-independently} issues two Nyms $N_1$ and $N_2$,
via a random 1-to-1 permutation---%
always giving each user exactly one Nym.
In the first communication round,
$N_1$'s owner announces a weak policy with a minimum buddy set size of 1,
but $N_2$ demands a buddy set size of 2.
The adversary later sees a non-null post to $N_1$ while user $B$ is offline,
and as $N_1$'s weak policy permits,
infers that $A$ must own $N_1$.
If each user owns exactly one Nym,
then the adversary can {\em also} infer that $B$ must own $N_2$,
violating $N_2$'s stronger policy.
With independent Nym assignment, in contrast,
the knowledge that $A$ owns $N_1$
gives the adversary no information about which user owns $N_2$,
because it is just as likely that $A$ owns {\em both} $N_1$ and $N_2$,
as it is that each user owns exactly one Nym.

\section{Buddies in Practical Systems}
\label{sec:design}

Since \apps conceptual model in Section~\ref{sec:model}
is unrealistically simple in several ways,
we now address key challenges of implementing \app
in practical anonymity systems.
For concreteness we will focus on the design of our \app prototype
built as an extension to
Dissent~\cite{corrigangibbs10dissent,wolinsky12dissent,corrigangibbs13verif},
but we also discuss the \app architecture's potential applicability
to other existing anonymous communication systems.

\xxx{ still to address:
- scheduling nuances
- mixing long-lived and short-lived pseudonyms
Actually the new challenge to me seems to be we need one medium to introduce Nyms
and another in which they can be used --
the following text makes it seem like we could just present a new Nym at random
}

\subsection{Decentralizing the Anonymizer}
\label{sec:design-decentral}

So far we have treated the Anonymizer as a trusted
``black box'' component,
but in a practical anonymity system
we do not wish to require users to trust any single component.
In a practical design, therefore, we replace \apps Anonymizer
with one of the standard decentralized schemes 
for anonymous message transmission,
such as mix-nets~\cite{chaum81untraceable,danezis03mixminion,leblond13anon},
DC-nets~\cite{chaum88dining,waidner89dining},
or verifiable shuffles~\cite{neff01verifiable,furukawa01efficient,brickell06efficient}.

While agnostic in principle to
specific anonymous communication mechanisms,
\app makes two important assumptions about the Anonymizer's design.
First, \app assumes the Anonymizer is already resistant
to basic, {\em short-term} network-level traffic analysis and timing
attacks~\cite{levine04timing,murdoch05low,shmatikov06timing}.
Without basic traffic analysis protection
necessary for {\em unlinkable} message traffic,
we cannot expect to achieve reliable long-term protection
for {\em linkable} posts via pseudonyms.
Second, we assume
the Anonymizer distributes trust across some group of servers,
and that {\em there exists} at least one trustworthy server in this group---%
although the user need not know {\em which} server is trustworthy.
This {\em anytrust} model~\cite{wolinsky12scalable} is already embodied in
the relays used in mix-nets or Tor circuits~\cite{dingledine04tor},
the ``authorities'' assumed in verifiable shuffles~\cite{neff01verifiable},
or Dissent's multi-provider cloud model~\cite{wolinsky12dissent}.

While ``ad hoc'' mix-nets and onion routing schemes are vulnerable to
many traffic analysis and active attacks,
variants such as
MIX cascades~\cite{pfitzmann91isdn,berthold00cascade,brickell06efficient} and
verifiable shuffles~\cite{neff01verifiable,furukawa01efficient}
offer formally provable traffic analysis resistance.
These systems typically work in synchronous rounds,
where users submit onion-encrypted ciphertexts
to a common set of mixes,
who serially decrypt and permute the ciphertexts
to reveal the anonymous plaintexts.
To resist traffic analysis,
users with no useful message to send in a round
must submit encrypted ``empty'' messages as cover traffic.

DC-nets~\cite{chaum88dining,waidner89dining}
similarly operate in synchronous rounds,
but derive anonymity from
{\em parallel} information coding techniques
rather than {\em serial} mixing,
achieving traffic analysis protection in fewer communication hops.
Dissent~\cite{corrigangibbs10dissent,wolinsky12dissent,corrigangibbs13verif}
adapts DC-nets to a practical and scalable client-server model.
By leveraging the parallel communication structure of DC-nets,
Dissent achieves per-round latencies orders of magnitudes lower
than a verifiable shuffle or cascade mix guaranteeing equivalent security.
Dissent thus forms a natural foundation
for our \app prototype to build on.

\subsection{Creating and Extending Nyms}
\label{sec:design-indep}

\app can conveniently represent
Nyms as public/private key pairs,
so that anyone may learn the public keys of all Nyms in existence,
but only a Nym's owner holds the corresponding private key.
Dissent runs Neff's verifiable key-shuffle~\cite{neff01verifiable},
once per communication {\em epoch},
to generate a list of public keys forming
a DC-nets transmission schedule for that epoch.
Each client submits one public key to the shuffle,
which the servers re-encrypt and randomly permute,
to produce a well-known list of {\em slot} keys.
The client holding a slot's matching key can identify its own slot,
but neither the servers nor {\em other} clients learn
who owns any honest client's slot.
Adapting this mechanism to create fresh Nyms in \app
introduces two further technical challenges:
assigning Nyms to users {\em independently} at random,
and enabling Nyms to have unlimited lifetimes.

\paragraph{Creating Nyms via Lotteries}

A simple way to create independent Nyms
meeting the requirements in Section~\ref{sec:policy-varying}
is by ``lottery.''
Each user submits a fresh public key to a verifiable shuffle,
the servers jointly pick one re-encrypted key at random
from the shuffled output to be a new Nym,
and discard all other keys.
For efficiency, we would prefer to generate fresh Nyms in batches,
amortizing the cost of the shuffle across lotteries for many Nyms,
while ensuring that each Nym is assigned independently.
One conceivable approach is for the servers to mint
a batch of e-cash ``coins''~\cite{chaum90untraceable},
encrypt each coin to a random key chosen from the shuffle's output,
and finally run a verifiable DC-nets
round~\cite{golle04dining,corrigangibbs13verif},
with one slot per coin, allowing each coin's ``winner''
to spend the coin and publish a fresh public Nym key.
We leave detailed exploration of this challenge to future work.

\com{
To guarantee Nym independence,
users of any Anonymizer could participate
in a verifiable shuffle
with the first entry in the final permutation
becoming the announced Nym for that iteration.
Alternatively as a tradeoff between practical efficiency and full independence,
assuming $n$ users insert $k \times m$ keys into a shuffle,
each of the $m - 1$ servers would throw away the last $n$ keys.
As a result each user obtains about $k$ Nyms.
Even after identifying $k$ Nyms as belonging to user $A$,
the adversary cannot infer with certainty that $A$ does not
{\em also} own some other Nyms.
Nevertheless, the resulting Nyms assignments are not {\em fully} independent.
Also we assume the resulting permutation from each shuffle operation
should be deterministic---lexicographical order, for example---%
so that no shuffler can influence the set of Nyms announced.
\xxx{	I still don't get why this approach is attractive -
	it sounds horribly inefficient (k * m * n items to shuffle!),
	and even when implemented correctly with constrained permutations,
	it doesn't really achieve nym independence!	}
}

\paragraph{Extending Nyms Across Epochs}

To enable Nyms to persist beyond one epoch,
clients can use each winning ticket from a Nym lottery
either to publish a fresh Nym key,
or to re-publish an old Nym key,
effectively ``reviving'' the old Nym 
and giving it a transmission slot in the new epoch.
A lottery winner might even publish {\em another} user's public Nym key,
effectively delegating the winning ticket's share of bandwidth in the new epoch
an arbitrary Nym whose content the lottery winner finds interesting.

When a client revives its own Nym in a new epoch,
the client must ensure that the set of users participating in the Nym lottery
is consistent with the existing Nym's anonymity policy:
e.g., that the user's buddies are also online.
If some participants go offline {\em during} a Nym lottery,
to avoid policy-checking races the servers must restart the lottery,
enabling clients to re-check the new participant set
before exposing an old Nym in the new epoch.



\subsection{Implementing the Policy Oracle}
\label{sec:design-policy}

Implementing the Policy Oracle as a single independent server
would require that all clients trust the Policy Oracle server
to implement their requested attack mitigation policies correctly.
Although a bad Policy Oracle server cannot {\em directly} de-anonymize users
since it does not know which users own each Nym,
it could---by intent or negligence---simply
make intersection attacks easy.
For this reason, \app leverages the {\em anytrust} server model
that the underlying Anonymizer already uses,
running a virtual replica of the Policy Oracle in ``lock-step''
on each of the anonymization servers.
The servers use standard distributed accountability
techniques~\cite{haeberlen07peerreview}
to cross-check each others' computation of Policy Oracle decisions,
halting progress and raising an alarm
if any server deviates from an agreed-upon deterministic algorithm
implementing the Policy Oracle.
These techniques apply readily to the Policy Oracle
precisely because it architecturally has access to no sensitive state,
and hence all of its state may be safely replicated.

\paragraph{Identifying Users to the Policy Oracle}
\label{sec:design-ident}

\apps Anonymizer shares the 
set of users currently online in each round with the Policy Oracle,
which implies that we must treat these online sets as ``public information''
that the adversary may also obtain.
As discussed in Section~\ref{sec:model-active}, however,
revealing {\em actual} user identities or locators for this purpose,
such as users' public keys or IP addresses,
risks {\em strengthening} a weak adversary into an ``omniscient'' adversary
for intersection attack purposes.

\app addresses this problem by permitting clients
to authenticate with the servers
via {\em linkable ring signatures}~\cite{liu04linkable,fujisaki07traceable}.
To connect,
each client generates a cryptographic proof that it holds the private key
corresponding to one of a {\em ring} of public keys,
without revealing {\em which} key the client holds.
In addition, the client generates and proves the correctness
of a {\em linkage tag},
which has a 1-to-1 relationship with the client's private key,
but	
is cryptographically unlinkable to any of the public keys
without knowledge of the corresponding private keys.
The servers track which clients are online via their linkage tags,
and provide only the list of online tags to the Policy Oracle in each round,
so the Policy Oracle can simulate an adversary's intersection attacks
without knowing which {\em actual} users are online each round.

Of course, the server that a client connects to directly
can associate the client's network-level IP address with its linkage tag,
and a compromised server may share this information with an adversary.
This linkage information does not help a global passive adversary,
who by definition obtains all the same information
the Policy Oracle obtains merely by monitoring the network,
but may help weaker adversaries perform intersection attacks
against those users who connect via compromised servers.
We see this risk as equivalent to the risk clients run
of connecting to a compromised ``entry relay''
in existing anonymity systems~\cite{danezis03mixminion,dingledine04tor}.
Compromised servers are just one of the many avenues
through which we must assume an adversary might monitor the network.

\subsection{Malicious Users and Sybil Attacks}
\label{sec:malicious}

While \app can measure, and optionally enforce a lower bound on,
the number of users comprising a Nym's \possy or indinymity set,
\app cannot guarantee that all those users
are providing {\em useful anonymity}.
In particular, if the owner of a Nym $N$
has specified a policy mandating a minimum buddy-set size of $K$,
but up to $F$ other clients may be colluding with the adversary,
then $N$'s owner may have to assume that its {\em actual}
minimum anonymity set size may be as little as $K-F$,
if all $F$ bad clients happen to---or somehow arrange to---%
land in the same buddy set as $N$'s owner.
Since in practical systems we don't expect users to have a reliable way
of ``knowing'' how many other clients are conspiring against them,
we treat $F$ as an unknown variable that users may simply have to ``guess''
and factor into their choices of possinymity or indinymity lower-bounds.
In this respect \app is no different from any other anonymity system
some of whose users may be compromised.

Reducing vulnerability to malicious clients may be
an argument in favor of random buddy-set formation
(Section~\ref{sec:model-buddy-sacrifice}).
Randomized policies may offer some guarantee that
the malicious users present in a Nym's initial user set
become ``evenly distributed'' among buddy-sets.
In any preferential, ``rep\-u\-ta\-tion-based'' formation scheme,
if the attacker can learn or correctly guess
the general ``level of reliability'' of a Nym $N$'s true owner---%
which may well be inferable from $N$'s posting record---%
then the attacker's compromised nodes might deliberately exhibit
a similar level of reliability in hopes of getting clustered together
in the owner's buddy set.
For such attacks to succeed, however,
the malicious users must be present at $N$'s creation
in order to fall in $N$'s possinymity set in the first place,
and the attacker must adjust their reliability profile
{\em after} learning ``enough'' about $N$'s owner, but
{\em before} too many buddy set splits have already occurred for $N$.
Thus, if the Policy Oracle builds up user reputation information
in a relatively conservative, long-term fashion
across the users' entire histories (e.g., from before $N$ appeared),
this may make it difficult for an attacker to ``steer''
malicious users' reliability profiles ``late in the game''
to implement a cluster attack $N$.
Clustering attacks nevertheless present a risk
that more randomized buddy set formation policies may reduce.

As in any distributed system,
an attacker may be able to amplify the effective numbers of malicious clients
via Sybil attacks~\cite{douceur02sybil},
creating many fake user identities.
\app addresses this risk by requiring users to be authenticated---%
via linkable ring signatures as detailed above---%
as owners of ``real'' identities in some Sybil attack resistant identity space.
\app is agnostic as to the exact nature of this public identity space
or how it is made resistant to Sybil attacks.
The current prototype simply is defined for ``closed'' groups,
defined by a static {\em roster} of public keys listing all members,
so the group is exactly as Sybil attack resistant as
whatever method the group's creater uses to form the roster.
To support open-ended groups,
\app could build on one of the many Sybil attack resistance schemes,
such as those based on social networks~\cite{tran09sybil,yu08sybillimit}---%
or could simply rate-limit Sybil attacks via some ``barrier to entry,''
e.g., requiring users to solve a CAPTCHA or receive a phone callback
to ``register'' an unknown public key for participation.

\section{Evaluation}
\label{sec:eval}

\begin{figure*}[t]
\centering
\begin{tabularx}{\linewidth}{XX}
\includegraphics[width=0.45\textwidth]{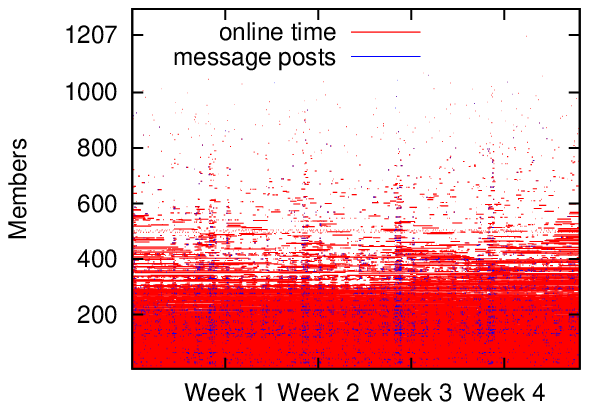}	&
\includegraphics[width=0.45\textwidth]{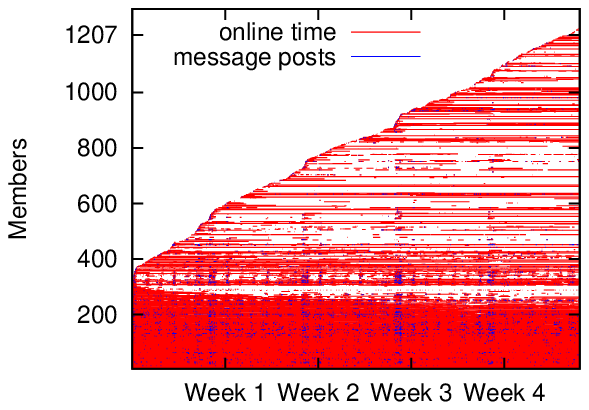}	\\
{\bf (a)} Members sorted by total time online			&
{\bf (b)} Members sorted by first appearance online		\\
\end{tabularx}
\caption{Visualization of user online periods over one month
	in EFnet's {\bf football} IRC chatroom.}
\label{fig:lines_online}
\end{figure*}

We now evaluate \apps utility,
using data we collected
from popular public IRC (Internet Relay Chat) chat rooms on EFnet servers.
After introducing our data collection and simulation approach,
we first explore ``ideal'' metrics quantifying levels of anonymity
achievable in principle under given conditions---%
metrics that depend {\em only} on the user behavior dataset
and not on any particular \app policy or loss mitigation algorithm.
Next, we apply these traces to an event-based \app simulator
to evaluate more realistic policies
against these ideals.
We consider na\"ive anonymous posting without \app,
then posting under policies that enforce minimum buddy-set sizes,
and policies that attempt to maximize possinymity.
Finally, we analyze the overheads \app induces
in the context of Dissent.

\subsection{Datasets and Simulation Methodology}

\xxx{	Better cite Clay's paper that does HTTP-based user studies}
To evaluate \apps utility,
we use traces taken
from popular public IRC (Internet Relay Chat) chat rooms on EFnet servers.
Unlike web traffic~\cite{wright08passive},
IRC logs record participants online status
in addition to activity or transmission of messages.
The online status becomes critical for systems
in which inactive but online users submit cover traffic,
such as \app assumes.
While BitTorrent traffic~\cite{bauer05bittorrent}
supports similar conventions as IRC,
online status and activity,
user behavior focuses transferring data between peers,
which does not make a strong correlation to the need
for long term intersection resistant Nyms.
Furthermore,
the buddy system focuses on anonymous group communication systems
that reveal all anonymous cleartexts to all users
with at least all servers privy to the input;
such behavior maps better to IRC and BitTorrent traffic than to Web traffic.

We monitored 100 (a limitation imposed by EFnet)
of the most active EFnet-based IRC rooms,
for over a month dating from November 26th through December 30th, 2012.
For each room, we obtained the following for each member:
joins, leaves, nickname changes, and messages.
Anecdotally, we found that users often temporarily disconnect from IRC
without IRC recognizing the disconnection.
This creates a period of time in which the user must use a secondary nickname,
then switch back to the original nickname once IRC recognizes the disconnect.
Unfortunately we have no statistics on average disconnection time,
but we were able to identify these scenarios and ``fix'' the data
such that the affected users appear to be continuously online.

Figure~\ref{fig:lines_online} visually illustrates the trace
collected from one sample IRC room, {\bf football},
plotting the time period a given user was online as a horizontal line.
Figure~\ref{fig:lines_online}(a)
sorts the 1207 unique users observed during the trace
vertically by total time the user was online during the observation period.
This graph shows that the online forum has
a ``core'' set of around 300 users who stay online most of the time,
while the remaining users come online for shorter periods at varying times,
with denser vertical stripes showing periodic patterns (e.g., football games).
\xxx{	weekly patterns?  Monday night football?  be more specific if possible.}
Figure~\ref{fig:lines_online}(b)
shows the same data with users sorted by {\em first} appearance online:
again about 300 users were online already at the start of the trace,
while new users appear subsequently at a fairly constant rate---%
with a fraction of these ``late arrivals'' remaining online
for the remaining trace period.

\com{	Subsumed into fig:ideal-anon
\begin{figure}[t]
\centering
\includegraphics[width=0.45\textwidth]{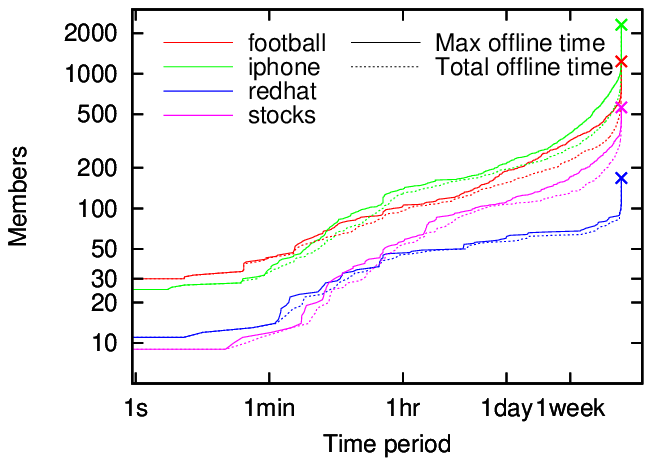}
\caption{Maximum user offline time for various IRC rooms}
\label{fig:max_off}
\end{figure}
}

\com{	I think max_off presents strictly more useful information than this.
\begin{figure}[t]
\centering
\includegraphics[width=0.40\textwidth]{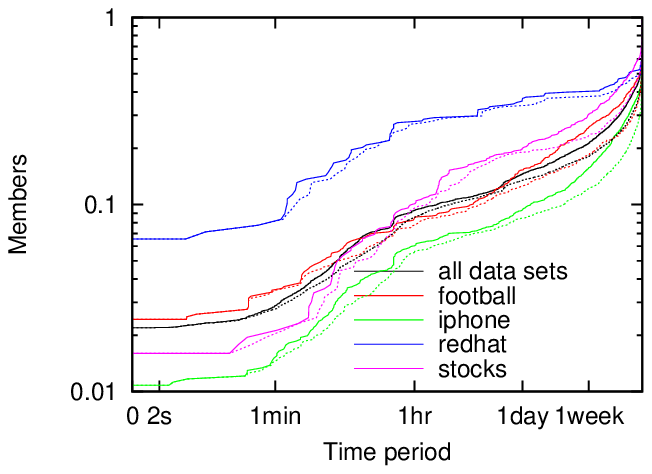}
\caption{Maximum user offline time for various IRC rooms
normalized for members.}
\label{fig:max_off_cdf}
\end{figure}
}

\com{Not needed any more, becomes redundant with fig:msgs
\begin{figure}[t]
\centering
\includegraphics[width=0.40\textwidth]{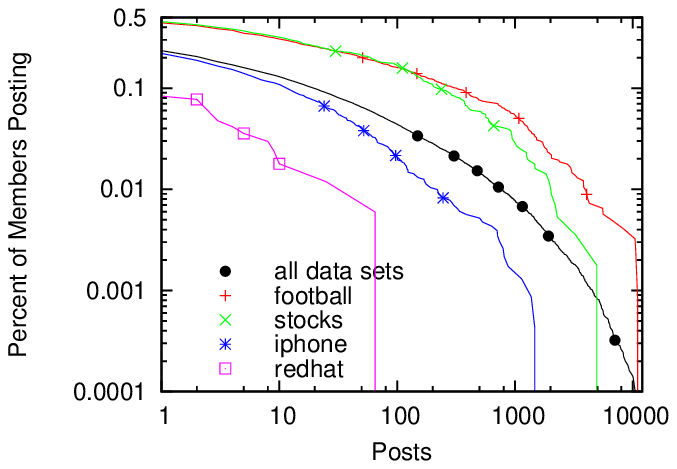}
\caption{Number of posts per user across IRC rooms}
\label{fig:posts_per_member}
\end{figure}
}

To analyze this data
we implemented an event-driven \app simulator in Python,
cleverly called the Anonymity Simulator (AS).
AS plays the role of users, the Anonymizer, and the Policy Oracle.
As input the AS takes an IRC trace,
time between rounds,
system-wide buddy and possinymity set sizes,
and buddy set formation policies.
We primarily focus on random buddy set formation policies and user online times.
\xxx{	what does it mean to "focus on...user online times"?	}
For the latter,
the AS can either use an initial period of the trace, a bootstrap period,
or use a deity mode and review the entire data set.
To better compare apples-to-apples,
random, bootstrapped, and deity mode evaluations all began
at the same time in the trace.
\xxx{	some of this is probably too low-level details for here,
	and maybe should be moved to the discussion of relevant experiments.}

\subsection{Ideal Anonymity Analysis}

\begin{figure*}[t]
\centering
\begin{tabularx}{\linewidth}{XX}
\includegraphics[width=0.45\textwidth]{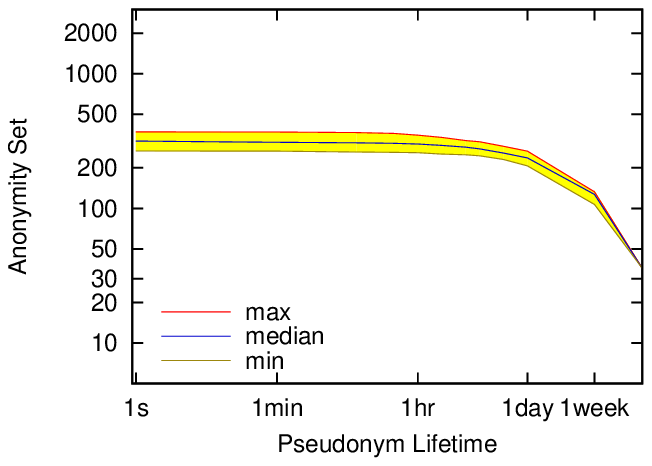}	&
\includegraphics[width=0.45\textwidth]{figs/max.eps}		\\
{\bf (a)} Ideal anonymity set size potentially achievable
	by a {\em low-latency} pseudonym of varying lifetime.
	The pseudonym can tolerate
	{\em no} offline time in members of its anonymity set.	&
{\bf (b)} Ideal anonymity potentially achievable for {\em long-lived}
	(1-month) pseudonyms,
	tolerating anonymity set members who go offline
	for a varying maximum offline time period.		\\
\end{tabularx}
\caption{Analysis of {\em ideal anonymity} potentially achievable
	based on user behavior in the IRC {\bf football} room.}
\label{fig:ideal-anon}
\end{figure*}

\com{	Subsumed into fig:ideal-anon
\begin{figure}[t]
\centering
\includegraphics[width=0.40\textwidth]{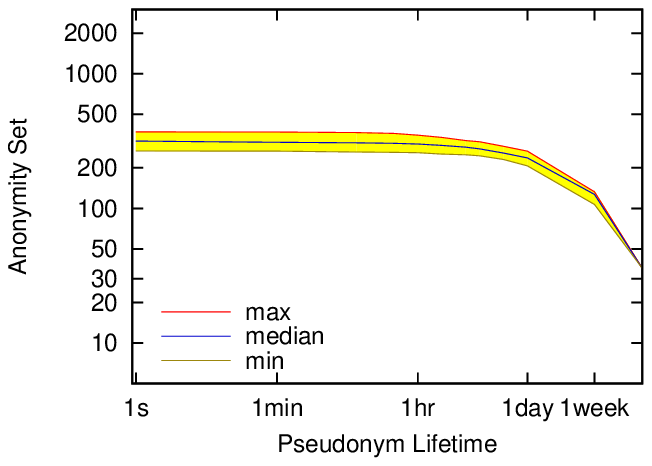}
\caption{Anonymity degradation over the lifetime of a low-latency pseudonym}
\label{fig:lifetime}
\end{figure}
}

We first use our IRC traces to explore upper bounds on the anonymity
we expect to be achievable in 
{\em any} system resistant to intersection attacks,
under a trace-driven scenario,
but independent of particular anonymity mechanisms or policies.
These experiments depend {\em only} on analysis of the IRC data,
and do not depend on \apps design or the \app simulator.
This analysis serves to deepen our understanding of user behavior
in realistic online forums,
and to establish realistic expectations of what
a system such as \app might achieve in principle.
We first consider anonymity potentially achievable
for low-latency communication using pseudonyms of varying lifetimes,
then focus on {\em long-lived} pseudonyms
in communication scenarios that can tolerate varying offline times
in members of anonymity sets.

\paragraph{Low-latency pseudonyms of varying lifetime}

We focus on the {\bf football} dataset,
considering all online periods of all 1207 users appearing in the trace
as visualized in Figure~\ref{fig:lines_online}.
We treat each contiguous online period lasting at least time $x$
as representing a pseudonym with lifetime $x$,
and compute an ideal anonymity set for that pseudonym
as the total number of users {\em also contiguously online}
during that pseudonym's lifetime.
This analysis pessimistically eliminates from the anonymity set
users with {\em any} offline period, however brief,
during the pseudonym's lifetime.

Figure~\ref{fig:ideal-anon}(a) summarizes
the distribution of these ideal anonymity set sizes,
for pseudonym lifetimes varying on the log-scale $x$-axis.
Pseudonyms used for up to about one hour
reliably achieve anonymity sets of at least 250 members,
and sometimes up to 375 members---%
between 20\% and 30\% of the total user population observed---%
suggesting that substantial resistance to intersection attack
may be achievable in large forums for short-lived pseudonyms.
\xxx{	David: double-check and perhaps fine-tune these numbers please.}
Achievable anonymity under these assumptions
falls off rapidly as pseudonym lifetime increases further,
however.
\com{
As a precursor to evaluating \app,
we consider the useful life time of a Nym and its
buddy set independent of other Nyms and buddy sets,
in other words,
the anonymity set of each Nym from the time it was first
used until the end of the data set.
For this analysis, we used the football data set
and plot various percentiles in Figure~\ref{fig:lifetime}.
After 1 hour, nearly all Nyms retain
the same level of anonymity as they began
and while there exists a sharper curve thereafter,
there still remains over 50\% of the anonymity set after 1 day,
25\% after 1 week,
and 12.5\% after 1 month.
}

\paragraph{Long-lived pseudonyms tolerant of offline periods}
Applications that demand truly long-lived pseudonyms,
such as blogs,
can often tolerate longer communication latencies.
Many anonymous communication schemes can {\em aggregate} user behavior
into selectable time periods,
such as batches in mix-nets~\cite{chaum81untraceable,danezis03mixminion},
or dropoff windows in DC-net systems
such as BlogDrop~\cite{corrigangibbs12scavenging}
or Verdict~\cite{corrigangibbs13verif}.
Users who come online and participate at {\em any time} during each window
are indistinguishable to the adversary,
enabling the system to tolerate users who go offline briefly
for periods shorter than that window.
\app policies can similarly be configured to tolerate
users who go offline for brief periods without anonymity loss,
as discussed in Section~\ref{sec:policy-poss}.
In all these cases, the main cost is to increase communication latency
up to that offline tolerance period.
Without focusing on any particular mechanism,
we now parameterize our ideal anonymity analysis on the assumption
that a long-lived pseudonym has {\em some} way to tolerate
users who go offline for periods up to a varying {\em maximum offline time}.

Figure~\ref{fig:ideal-anon}(b) shows the ideal anonymity achievable
for a hypothetical long-lived pseudonym whose lifetime is
the {\em entire} 1-month observation period of the IRC trace,
assuming that pseudonym can tolerate a varying maximum offline time
shown on the log-scale $x$-axis.
Since this scenario gives us only one ``data point'' per IRC trace
for a given maximum offline period,
the figure shows the same analysis for several different IRC datasets.
Consistent with the right edge of Figure~\ref{fig:ideal-anon}(a),
long-lived pseudonyms that cannot tolerate even short offline periods---%
as required for low-latency communication---%
achieve small anonymity sets consisting of
only the 10--30 users in each trace who were
continuously throughout the trace.
(The leftmost portion of this graph may be optimistic, in fact,
as the IRC server may not have logged temporary network disconnections
shorter than the TCP timeout of around 30 seconds.)
A long-lived \pnym that can tolerate disconnections up to one hour,
however,
can achieve 100-user anonymity sets across the 1-month trace
in the {\bf football} and {\bf iphone} rooms.
A \pnym that can tolerate 1-day disconnections---%
realistic for a blog whose author posts at most once per day anyway---%
can achieve still larger anonymity sets of around 200 users,
or 6\% of the {\bf football} group's total observed membership.

\com{
Buddy sets obtain the most utility
when bundling buddies with similar behaviors together,
namely, offline time.
Our interest in maximum offline time as opposed to
total offline time reflects \apps need for users to be consistently online
for extended periods of time,
to maintain anonymity under intersection attacks.
In other words,
Buddies can participate in every round within the trace
if the round intervals are equal to or greater than their maximum offline time.
Figure~\ref{fig:max_off} examines total offline time
versus maximum offline time across the entire data set.
While not significantly different,
we do obtain more members when considering maximum offline time.
The number of users in each trace was:
football 1207, iphone 2290, redhat 160, and stocks 558.
}

In general, the IRC datasets exhibit a common pattern
where a small set of dedicated users is almost always online,
a larger set--roughly 15\% to 20\%--who
show up about once an hour to once a day,
and a large set of ephemeral users
comprise around 80\% of the total population.
We do not expect ephemeral users to contribute usefully
to the anonymity sets of long-lived Nyms,
but it is important for \app to operate in their presence,
and ephemeral users can increase anonymity
for short-lived Nyms as Figure~\ref{fig:ideal-anon}(a) shows.


\subsection{Possinymity Analysis and Enforcement}

\begin{figure}[t]
\centering
\includegraphics[width=0.47\textwidth]{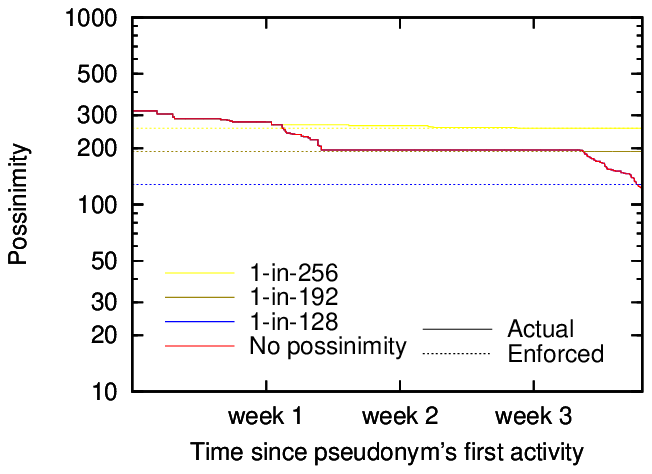}
\caption{Evolution of worst-case possinymity over trace
	\vspace{1em}}
\label{fig:possinymity}
\end{figure}

We now explore the behavior of Buddies' possibilistic \anon metric,
as defined in Section~\ref{sec:model-poss},
under our IRC trace workloads.
While the above ideal \anon analysis generically explored
the levels of \anon a ``hypothetical'' pseudonym might achieve,
we now wish to use our traces to model {\em specific} pseudonyms
and explore their behavior under specific Buddies policies.
To do so, we consider the messages posted by each IRC user
to represent one pseudonym in our model,
we take the times these IRC messages appeared to represent a ``schedule''
of the times the modeled pseudonym owner had a message to post,
and we define the nominal {\em lifetime} of each modeled pseudonym
as the time from the first message to the last message.
We then replay this activity under AS, the Anonymity Simulator,
to evaluate the behavior of Buddies' metrics under these activity traces,
and the effect of Buddies policies on {\em actual} communication behavior,
taking into account the (sometimes indefinite) delays
that Buddies sometimes imposes to preserve \anon.

Figure~\ref{fig:possinymity} shows how \possy evolves
across the lifetimes of pseudonyms with 1-second round times
(no offline time tolerance).
The $x$-axis represents time since since each pseudonym's inception---%
i.e., from its first scheduled message.
The $y$-axis value of the solid lines show
minimum or {\em worst-case} \possy across
all pseudonyms whose lifetimes extend at least to that point.
Each line color in the graph reflects behavior under Buddies policy
that enforces a different \possy lower bound,
whose enforced level is represented by the corresponding horizontal dotted line.

Consistent with our expectations from Figures~\ref{fig:lines_online}
and~\ref{fig:ideal-anon}(a),
\possy starts at the number of users online at time of first post,
and gradually decreases as users ``churn'' offline
and get intersected out of the pseudonym's \possy set
at a relatively constant rate.
In practice, enforcing a lower bound leaves pseudonyms' behavior unaffected
until just before a message post would decrease the pseudonym's \possy
below the lower bound.
At this point the \pnym becomes unable to post---%
i.e., Buddies deliberately compromises availability to preserve \anon---%
leaving \possy constant after that point in the trace.
\xxx{make sure this actually shows up in the graph, or else revise text.}

\subsection{Indinymity Analysis and Buddy Sets}

\xxx{standardize on buddy-set or buddy set everywhere}

\begin{figure*}[t]
\centering
\begin{tabular}{ccc}
\includegraphics[height=1.8in]{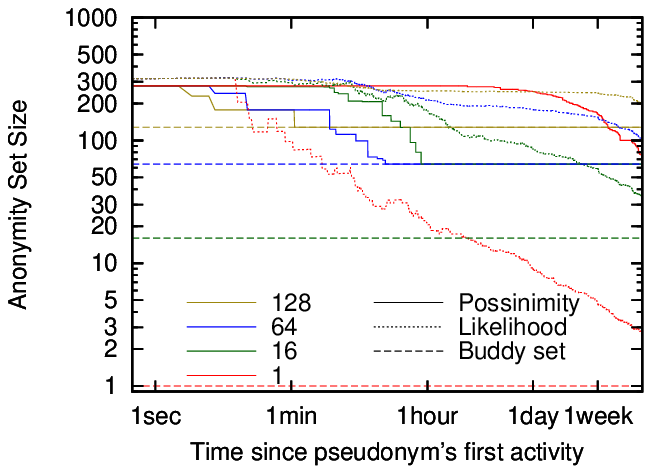}	&
\includegraphics[height=1.8in]{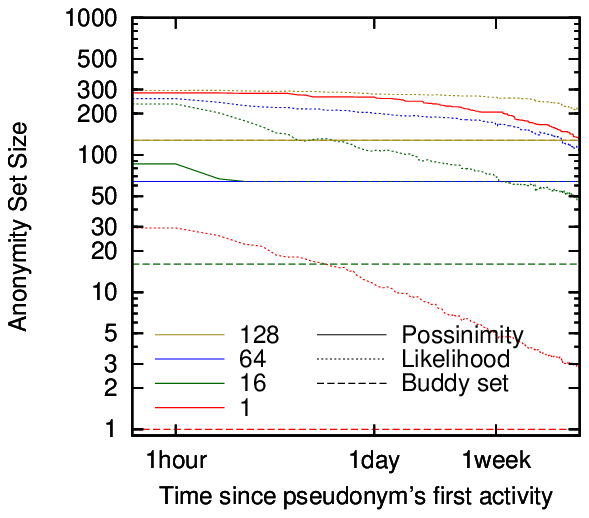}	&
\includegraphics[height=1.8in]{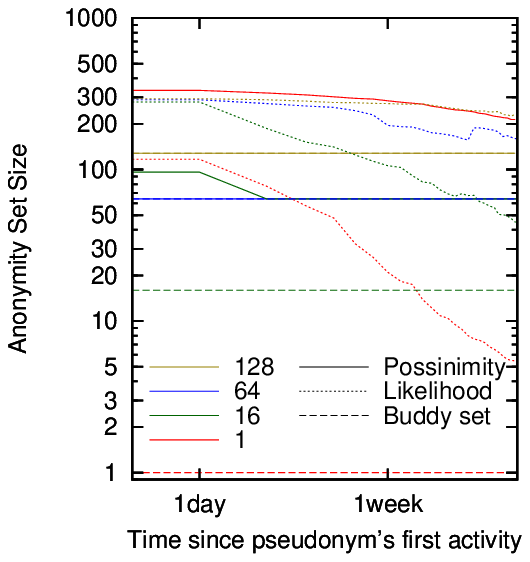}	\\
{\bf (a)} 1-second rounds					&
{\bf (b)} 1-hour rounds						&
{\bf (c)} 1-day rounds						\\
\end{tabular}
\caption{Measured worst-case anonymity loss over time
	under one example probabilistic intersection attack.}
\label{fig:indinymity}
\end{figure*}

We now explore probabilistic attacks of the form
discussed in Section~\ref{sec:model-prob},
and Buddies' ability to protect users against such attacks via buddy sets.
To validate the protection Buddies provides,
we model one {\em particular} probabilistic intersection attack,
in which the attacker observes the messages posted by a pseudonym,
assumes the owner decided whether to post in each round
with a fixed, independent probability $p$,
and uses the analysis outlined in Section~\ref{sec:model-prob}
to divide users into classes based on likeliness of owning the pseudonym.
The attacker then makes a ``best guess'' at which user owns that pseudonym,
and we compute the attacker's chance of guessing correctly:
$1/k$ if the owner is within the $k$-user equivalence class 
the attacker judges most likely, and $0$ if not.
We make no claim that this particular attack is representative
of the attacks most likely to be mounted by real-world adversaries,
but merely use it as an example
to test and visualize the eff\-ec\-tiveness of buddy sets.
We reiterate that Buddies' security is based ultimately
not on knowledge or expectations of particular attacks,
but on an indistinguishability principle
that applies to all probabilistic attacks
using observed online/offline behavior as inputs.

Figure~\ref{fig:indinymity} shows pseudonyms' measured worst-case \anon
under this probabilistic attack,
again as a function of the length of time each pseudonym has been active.
Each line color represents a different buddy set size enforcement policy,
with dotted lines representing effective \anon
against the probabilistic guessing attack,
solid lines representing measured possinymity for reference,
and dashed lines representing the buddy set size---%
and hence indinymity lower bound---%
enforced by the respective policy.
Note that these graphs have logarithmic $x$-axes,
unlike Figure~\ref{fig:possinymity}.
While possinymity decreases at a relatively constant rate with user churn,
the probabilistic attack is much faster---%
ef\-fec\-tive\-ly eliminating the vast majority of an unprotected user's
initial \anon set within the first hour since a pseudonym starts posting,
for example.
Nevertheless, for a given enforced buddy set size,
the probabilistic attack only ap\-proach\-es, but never violates,
the enforced indinymity bound.

\xxx{explain why graphs aren't monotonic?}

\subsection{Effect on Usable Pseudonym Lifetime}

\begin{figure*}[t]
\centering
\begin{tabular}{ccc}
\includegraphics[height=1.8in]{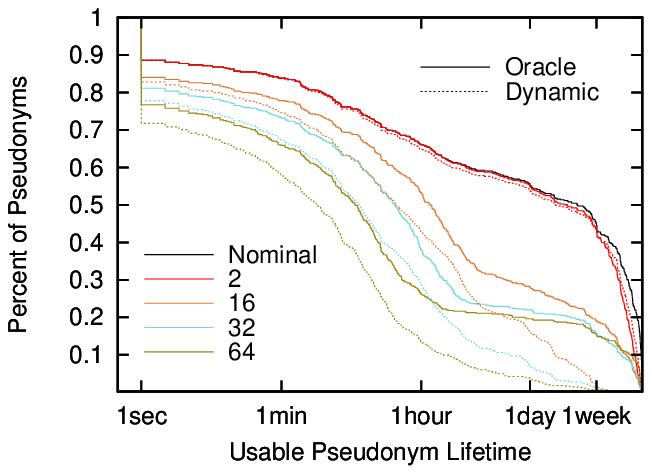}	&
\includegraphics[height=1.8in]{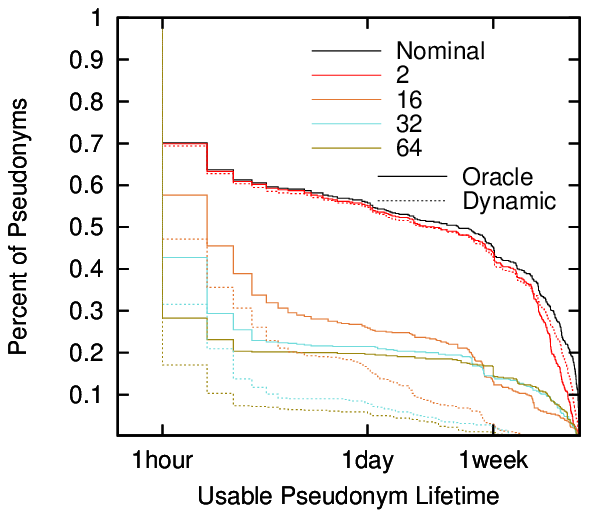}	&
\includegraphics[height=1.8in]{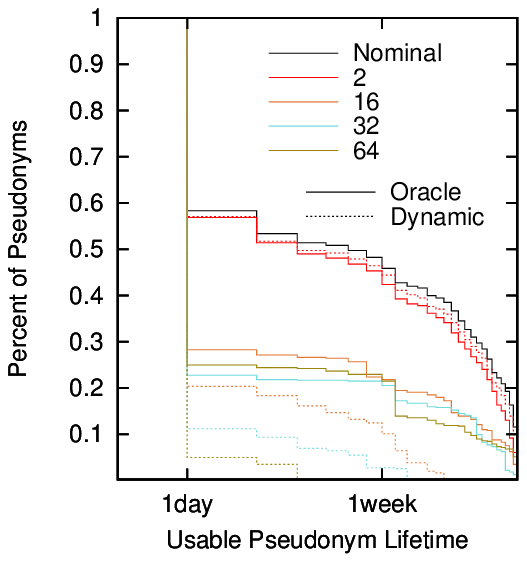} \\
{\bf (a)} 1-second rounds					&
{\bf (b)} 1-hour rounds						&
{\bf (c)} 1-day rounds						\\
\end{tabular}
\caption{CDFs showing distributions of pseudonym lifetimes,
	with and without indinymity enforcement via buddy sets.}
\label{fig:usability}
\end{figure*}

While the above experiments validate Buddies' effectiveness
at mitigating \anon loss through
both possibilistic and probabilistic intersection attacks,
this security necessarily comes with usability tradeoffs.
Buddies policies may make a pseudonym unusable for posting messages
before the owner naturally finishes using the pseudonym,
and in still usable pseudonyms, messages may be delayed.

To explore Buddies' effect on pseudonym usability,
Figure~\ref{fig:usability} contains CDFs showing the distributions
of {\em useful lifetimes} of pseudonyms
under various enforced buddy set sizes.
As the relevant comparison baseline,
the black line representing the {\em nominal} case
shows the distribution of pseudonym lifetimes in the IRC trace
without any Buddies policy.
We take this line to represent the distribution of time periods
during which traced users {\em intended} to use a pseudonym to post messages.
Each colored line in contrast shows the distribution
of {\em actual}, useful pseudonym lifetimes
upon enforcing a given enforced buddy set size.

For many pseudonyms, unfortunately,
resistance to probabilistic intersection attack comes at a high cost.
As Figure~\ref{fig:usability}(a) shows, for example,
50\% of the pseudonyms modeled in the trace
have nominal lifetimes of at least one day---%
meaning that their owners ``would like to'' use them for one day---%
but under enforced buddy set sizes of 32 or more,
50\% of pseudonyms remain usable only for about an hour under Buddies.
{\em Some} pseudonyms remain usable for much longer even under Buddies, however:
in particular, those pseudonyms whose owners are long-lived
and fall into the same buddy set with other long-lived, reliable users.
\xxx{adjust based on final graphs.}
\xxx{what about different round times?  anything to say? 
	do we even want to keep those graphs, given that the behavior
	seems pretty much the same for longer round times?}

For each buddy set size,
the graph additionally contrasts two schemes for
dividing users into buddy sets.
The {\em Oracle} scheme ``sorts'' users into buddy sets
based on their maximum offline time,
as measured for Figure~\ref{fig:ideal-anon}(b),
thereby clustering users with similar reliability levels together
from the perspective of an oracle who can ``see'' into the future.
The dotted lines, for comparison,
reflect a more realistic, {\em dynamic} scheme
designed along the lines discussed in Section~\ref{sec:policy-ind}.
This scheme clusters users into buddy sets
dynamically based only on {\em recent past} reliability history,
up to to the point in time when a large buddy set must be subdivided
in order to keep a pseudonym usable.
Since past reliability is not a perfect predictor of future reliability,
the realistic, dynamic scheme incurs some further loss
of effective pseudonym lifetime.
The current dynamic scheme is simplistic and can likely be improved,
so we expect the utility actually achievable in practical systems to lie
somewhere between the respective solid and dotted lines.

\subsection{Effect on Pseudonym Messaging Delay}

\begin{figure*}[t]
\centering
\begin{tabular}{ccc}
\includegraphics[height=1.8in]{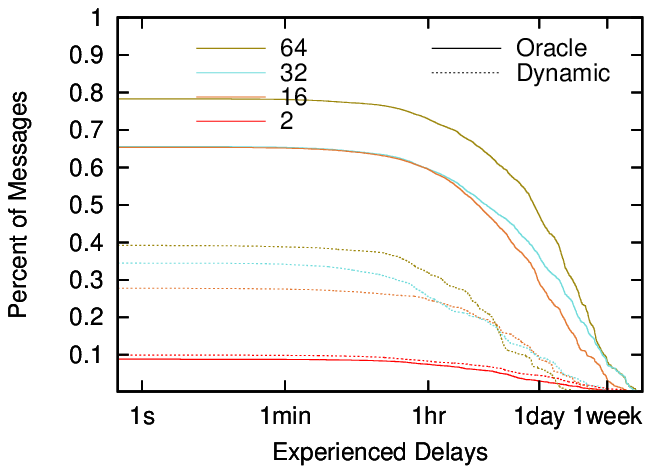}	&
\includegraphics[height=1.8in]{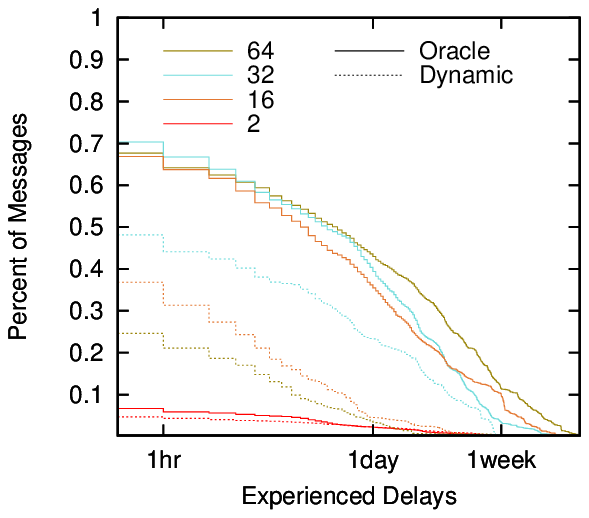}	&
\includegraphics[height=1.8in]{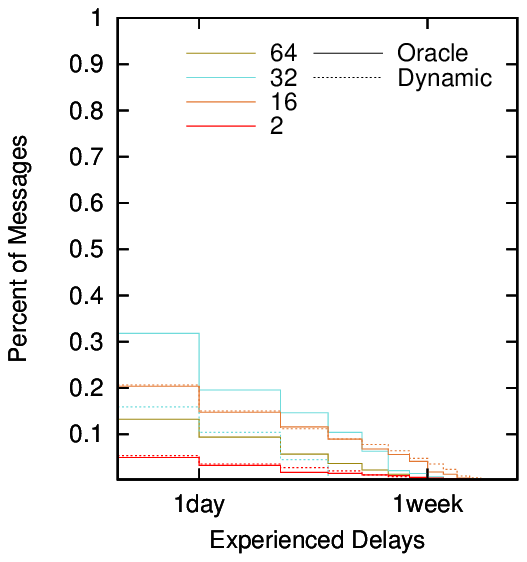} \\
{\bf (a)} 1-second rounds					&
{\bf (b)} 1-hour rounds						&
{\bf (c)} 1-day rounds						\\
\end{tabular}
\caption{CDFs showing distributions of delays imposed on
	successfully transmitted messages under Buddies.}
\label{fig:delay}
\end{figure*}

Finally, for messages that users successfully transmit via pseudonyms
under various enforced buddy set sizes,
Figure~\ref{fig:delay} shows the distribution of artificial {\em delays}
that Buddies imposes on those message transmissions
to preserve indinymity.
Unsurprisingly, the percentage of messages experiencing delays
increases with buddy set size,
since any pseudonym's owner must effectively wait to post
until all her buddies are also online,
and large buddy sets increase the likelihood of some buddy being offline
at the time of a desired message transmission.
When a message {\em is} delayed, unfortunately,
it is commonly delayed by at least one hour,
and under large buddy set sizes often by a day or more.
This result further confirms our intuition that
{\em long-term} resistance to probabilistic intersection attacks
in dynamic networks may really be feasible
only for delay-tolerant transmission.

We contrast the messaging delays experienced
under the two buddy set formation schemes above:
one based on an oracle's perspective of users' long-term reliability,
the other based on a more realistic dynamic algorithm.
In this case,
we find somewhat surprisingly that the dynamic scheme
actually delays substantially fewer messages
than the oracle scheme.
This is because the oracle scheme sorts buddies into sets
based only on {\em global} reliability measured over the entire trace,
and may inappropriately group together
users who were online for similar durations {\em but at different times}.
As the dynamic scheme builds buddy sets using recent information
localized in time,
it better groups ephemeral users who are online at similar times.
Neither of these buddy set formation schemes
are likely to be close to optimal,
and we consider improving them to be a useful area for future work.


\subsection{Practical Considerations}

The integration of \app into Dissent
sheds light into both the overheads of \app in a real system
and the implementation complexities that \app induces.
We built both \app and a web service for querying the various \app meters
and we modified Dissent~\cite{wolinsky12dissent} to
1) support the reactive and proactive analysis in the anonymity protocol and
2) transmit the set of online members (those with ciphertexts used in Dissent)
along with the cleartext messages at the end of each round.
\app totaled 528 lines of C++ code,
while Dissent incurred only 41 lines of additional C++ code.
Included within the 528 lines of C++ code,
\app comes with both a static and dynamic policy
each weighing in at 89 and 172 lines of code, respectively.

In Dissent, the buddy set concept applied cleanly;
however, in the current Dissent implementation,
the maintenance of possinymity is not ideal.
While DC-nets theoretically allow
serial processing of anonymous cleartexts
allowing finer grained control over possinymity,
currently, Dissent servers perform the cleartext revealing process
for all scheduled Nyms in parallel,
which limits possinymity evaluations
to occur only before processing the first anonymous cleartext.
Performing the operation iteratively would require adding
additional interaction among the servers,
potentially adding significant overhead.
Overhead would be negligible
only for rounds with interval time $t$
in which there were $m$ scheduled Nyms
with inter-server network latency of $l$,
where $m \times l \ll t$.
Fortunately, the Dissent implementation does support
scheduling Nyms during different intervals,
and therefore \app can still make
both possinymity and indinymity sets largely independent for different Nyms.

\begin{figure}[t]
\centering
\includegraphics[width=0.40\textwidth]{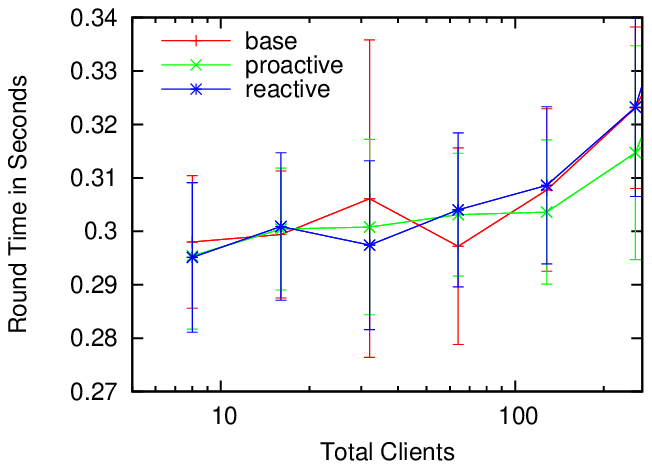}
\caption{The overhead of using \app in Dissent.}
\label{fig:dissent}
\end{figure}

Using our Dissent implementation of \app,
we focused evaluations on the additional delay added by \app.
We constructed a network of 8 server machines, 64 client machines,
and from 8 to 512 clients (running up to 8 clients on each client machine).
Clients have a 50 millisecond delay
and aggregate 100 M/bit connection to servers,
while servers have a 10 millisecond delay to other servers,
but each has a dedicated 100 M/bit connection to every other server.
To focus on the overheads \app incurs,
we avoided sending anonymous plaintexts across Dissent,
The results, shown in Figure~\ref{fig:dissent},
indicate that \app imposes negligible overhead over Dissent---%
within measurement error---%
because performance costs are dominated by the
expensive asymmetric cryptography used
in signing and verifying messages.
\xxx{	Let's get these results back in in some form!}

\section{Related Work}
\label{sec:related}

The utility of pseudonyms has been well-recognized since Chaum's seminal paper
on mix networks~\cite{chaum81untraceable}.
Pseudonymity has motivated much work in
anonymous authentication~\cite{lysyanskaya00pseudonym}
and signature schemes~\cite{liu04linkable,fujisaki07traceable}.
One way to
protect distributed systems from Sybil attacks~\cite{douceur02sybil}
is to build online pseudonyms 
atop ``real-world'' identities~\cite{ford08nyms}.
None of these approaches protect a pseudonym's owner
from being traced via network monitoring, however.
We concur with recent proposals
to integrate pseudonymity
into network architecture~\cite{han13expressive},
although this gateway-based proposal
unfortunate\-ly has the same ``single point of failure'' weakness
as a conventional single-hop
proxy or commercial VPN service~\cite{anonymizer}.

Multi-hop mix networks~\cite{danezis03mixminion}
and onion routing systems~\cite{dingledine04tor}
address this single point of failure,
but remain vulnerable to many traffic analysis attacks.
Pseudonymous communication is by now well-known to be highly vulnerable
to long-term intersection attacks~\cite{raymond00traffic,kedogan02limits},
later strengthened into statistical disclosure
attacks~\cite{mathewson04disclosure,danezis04statistical,wright08passive}.
Padding communications to a constant rate with
dummy traffic~\cite{dai96pipenet,berthold02dummy,freedman02tarzan,leblond13anon}
can slow---but not stop---passive
intersection attacks~\cite{mathewson04disclosure}.

Padding may not even slow {\em active} attacks, however,
where an attacker deliberately perturbs performance
to trace a circuit~\cite{murdoch05low,shmatikov06timing,evans09practical}.
Active attacks are eminently realistic,
being already widely used for
Internet censorship~\cite{verkamp12inferring,gill13characterizing},
for example.
Buddies is the only architecture we are aware of
that addresses both passive {\em and} active intersection attacks,
by ``collectivizing'' the anonymizer's control plane logic
into a Policy Oracle component that cannot see---and thus cannot leak---%
sensitive information (Section~\ref{sec:model}),
even when replicated for accountability (Section~\ref{sec:design-policy}).
This architecture ensures that Buddies' attack mitigation policies
apply regardless of whether clients churn ``normally,''
or are {\em forced} to slow or go offline
due to deliberate denial-of-service.

\com{
\app assumes that a user is included within the online roster
if and only if that user
has an active link to the Anonymizer
and sends appropriate traffic at schedule intervals.
As a parallel, \app could perhaps support dummy services
that send dummy traffic (null or cover traffic)
for an offline user
thus making that user appear online~\cite{berthold02dummy}.
While we have yet to make considerable effort in this direction,
we recognize several challenges.
\app makes conservative estimates about a user's anonymity
and also requires online state of participants.
By using a dummy service,
\app would lose this information
whereas an adversary may actually know the user's real online state
invalidating \app and likely making anonymity worse
for other participants.
}

The Java Anonymous Proxy~\cite{berthold00anonymity}
incorporates an ``Anonym-O-Meter''~\cite{federrath11monitor},
to give users an indication of their current anonymity level.
This meter does not address intersection attacks,
but serves as a precedent for \apps computation and reporting
of possinymity and indinymity metrics.
\com{	not sure what this is saying??? "related to the activity"?? -baf
The meter ignores intersection attacks, but does investigate
another potential interesting aspect the amount of anonymity
related to the activity of participants.
}

Prior systems also protect anonymity within well-defined groups.
Tarzan~\cite{freedman02tarzan} organizes an overlay of {\em mimics},
where each user maintains constant {\em cover} traffic
with $k$ other mimics to mitigate traffic analysis attacks.
Systems based on DC-nets~\cite{chaum88dining},
such as Herbivore~\cite{sirer04eluding}
and earlier versions of
Dissent~\cite{corrigangibbs10dissent,wolinsky12dissent,corrigangibbs13verif},
achieve provable traffic analysis resistance for {\em unlinkable} messages
in a {\em single} communication round.
Since every group member typically knows the online status of every other,
however,
linkable transmissions using pseudonyms can make such systems
{\em more} vulnerable to intersection attack
than ``amorphous'' systems such as Tor.
\app addresses this risk by using linkable ring signatures
to authenticate and ``tag'' users (Section~\ref{sec:design-ident}).

Hopper and Vasserman~\cite{hopper06on}
establish anonymity among sets of $k$ members in a mix,
similar to buddy sets,
and explore the resistance of these $k$-anonymity sets
to statistical disclosure attacks.
\app builds on this approach to offer users dynamic anonymity monitoring
and active controls over tradeoffs between anonymity and performance.
Aqua~\cite{leblond13anon}
uses padded, multipath onion routing to achieve efficiency
and reduce vulnerability to traffic analysis.
We expect \app be synergistic with designs like Aqua's,
by providing a stronger and more controllable notion
of intersection attack resistance
than currently provided by Aqua's $k$-sets.

\section{Conclusion}
\label{sec:conc}

\app offers the first systematic architecture
addressing long-term intersection attacks in anonymity systems,
by offering passive metrics of vulnerability
and active control policies.
While only a first step leaving many open questions,
our trace-based simulations and working prototype
suggest that \app may point to practical ways of
further protecting anonymity-sensitive users of online forums.

\subsection*{Acknowledgments}

We would like to thank
Joan Feigenbaum, Aaron Johnson, Peter Druschel, our shepherd Clay Shields,
and the anonymous reviewers
for their helpful feedback on this paper.
This material is based upon work supported by the Defense Advanced Research
Agency (DARPA) and SPAWAR Systems Center Pacific, Contract No.
N66001-11-C-4018.

\bibliographystyle{abbrv}
\bibliography{net,sec,os,soc}


\end{document}